\documentclass[10pt]{article}

\usepackage{microtype}
\usepackage{graphicx}
\usepackage{subfigure}
\usepackage{booktabs} 
\usepackage{csquotes}

\usepackage[hidelinks]{hyperref}
\usepackage{listings}
\usepackage{url}

\usepackage[mode=buildmissing]{standalone} 
\usepackage{tikz}
\usepackage{pgfplots}
\usepgfplotslibrary{groupplots,dateplot}
\usetikzlibrary{patterns,shapes.arrows}
\pgfplotsset{compat=newest}

\usepackage[preprint]{tmlr}

\usepackage{amsmath,amsfonts,bm}

\def\eqref#1{equation~\ref{#1}}

\def\1{\bm{1}}

\DeclareMathAlphabet{\mathsfit}{\encodingdefault}{\sfdefault}{m}{sl}
\SetMathAlphabet{\mathsfit}{bold}{\encodingdefault}{\sfdefault}{bx}{n}

\usepackage{amsmath}
\usepackage{amssymb}
\usepackage{mathtools}
\usepackage{amsthm}

\usepackage[capitalize,noabbrev]{cleveref}

\usepackage{acro}
\acsetup{make-links}

\DeclareAcronym{ml}{
    short = ML,
    long = machine learning,
}
\DeclareAcronym{plos}{
    short = PLoS,
    long = Public Library of Science,
}
\DeclareAcronym{icml}{
    short = ICML,
    long = International Conference on Machine Learning,
}
\DeclareAcronym{neurips}{
    short = NeurIPS,
    long = Conference on Neural Information Processing Systems,
}
\DeclareAcronym{pip}{
    short = PIP,
    long = package installer for Python,
}
\DeclareAcronym{pypi}{
    short = PyPI,
    long = Python Package Index,
}
\DeclareAcronym{pypa}{
    short = PyPA,
    long = Python Packaging Authority,
}
\DeclareAcronym{pmlr}{
    short = PMLR,
    long = Proceedings of Machine Learning Research
}
\DeclareAcronym{iclr}{
    short = ICLR,
    long = International Conference on Learning Representations
}
\DeclareAcronym{pep}{
    short = PEP,
    long = Python Enhancement Proposal
}
\DeclareAcronym{tmlr}{
    short = TMLR,
    long = Transactions on Machine Learning Research
}
\DeclareAcronym{mloss}{
    short = MLOSS,
    long = Machine Learning Open Source Software
}
\DeclareAcronym{aistats}{
    short = AISTATS,
    long= International Conference on Artificial Intelligence and Statistics
}
\DeclareAcronym{mlrc}{
    short = MLRC,
    long= Machine Learning Reproducibility Challenge
}
\DeclareAcronym{JIT}{
    short = JIT,
    long= just-in-time
}
\DeclareAcronym{pdf}{
    short = PDF,
    long = Portable Document Format
}
\DeclareAcronym{ci}{
    short = CI,
    long = Continuous Integration
}

\definecolor{tabBlue}{HTML}{1f77b4}
\definecolor{tabOrange}{HTML}{ff7f0e}
\definecolor{tabGreen}{HTML}{2ca02c}
\definecolor{tabRed}{HTML}{d62728}
\definecolor{tabPurple}{HTML}{9467bd}
\definecolor{tabBrown}{HTML}{8c564b}
\definecolor{tabPink}{HTML}{e377c2}
\definecolor{tabGray}{HTML}{7f7f7f}
\definecolor{tabOlive}{HTML}{bcbd22}
\definecolor{tabCyan}{HTML}{17becf}

\def\month{MM}  
\def\year{YYYY} 
\def\openreview{\url{https://openreview.net/forum?id=XXXX}} 

\title{More Rigorous Software Engineering Would Improve Reproducibility in Machine Learning Research}

\author{\name Moritz Wolter \email moritz.wolter@uni-bonn.de
\addr Bonn University, Germany
\AND
\name Lokesh Veeramacheneni \email lokiv@uni-bonn.de
\addr Bonn University, Germany
\AND
\name Charles Tapley Hoyt \email cthoyt@gmail.com
\addr RWTH Aachen University, Germany
}

\begin{document}

    \maketitle

    \begin{abstract}
        While experimental reproduction remains a pillar of the scientific method, we observe that the software best practices supporting the reproduction of \ac{ml} research are often undervalued or overlooked, leading both to poor reproducibility and damage to trust in the \ac{ml} community.
        We quantify these concerns by surveying the usage of software best practices in software repositories associated with publications at major \ac{ml} conferences and journals such as \acs{neurips}, \acs{icml}, \acs{iclr}, \acs{tmlr}, and \acs{mloss} within the last decade.
        We report the results of this survey that identify areas where software best practices are lacking and areas with potential for growth in the \ac{ml} community.
        Finally, we discuss the implications and present concrete recommendations on how we, as a community, can improve reproducibility in \ac{ml} research.
    \end{abstract}


    \section{Introduction} \label{submission}

    Scientific claims can only be considered empirical or scientific if we are capable of testing them~\citep{popper2005logic}.
    Theoretical papers are often self-contained, while numerical work typically depends on code.
    The large numerical part of \acf{ml} research, therefore, rests upon empirical foundations.
    Code reproducibility is a key concern in this branch, as it enables others to repeat and build upon prior work.

    This paper focuses on what we, as a community, are already doing to enable the reproducibility of our work and how we can improve further.
    Since machine learning heavily relies on software, we review established software engineering best practices, systematically estimate their adoption in source code repositories associated with publications in major \ac{ml} conferences and journals between 2018 and 2025, and ultimately argue for more rigorous software engineering in \ac{ml} research.
    Our assessment is based on data from a large-scale web crawl of repository links in conference papers.
    Using the crawled data, we identify current research trends and gaps in software engineering.
    Given the popularity of \ac{ml} frameworks like PyTorch~\citep{paszke2017automatic}, Jax~\citep{jax2018github}, and Tensorflow~\citep{tensorflow2015-whitepaper}, Python has become the most widely used language in \ac{ml}.
    Consequently, the \ac{neurips} guide for releasing code~\citep{NeuripsCodeguide} is written exclusively with Python in mind.
    Therefore, while our ideas for examining reproducibility are language-agnostic, our assessment focuses on Python-specific best practices.

    Finally, we provide actionable recommendations for how we as a community can improve further.
    We make the source code for our analysis available online
    at \url{https://github.com/BonnBytes/ml-swe-analysis}
    such that it can be periodically refreshed and later extended to cover additional conferences and journals.

    \section{Related work}\label{sec:related-work}

    Before reviewing software engineering best practices, we survey recent works from the \ac{ml} community on the hallmarks of reproducibility, how new ideas in reproducibility have been operationalized, the relationship between open source and reproducibility, and the effect of reproducibility on uptake and citation.

    \paragraph{Defining reproducibility}\cite{tatman2018practical} proposed a distinction between \ac{ml} reproduction and \ac{ml} replication in which reproduction refers to recreating the exact results reported by a paper, while replication describes the application of the described methods to another dataset.

    \paragraph{Hallmarks of reproducibility}\cite{hutson2018artificial} and \cite{haibe2020transparency} described how missing source code and data are key obstacles to \ac{ml} reproduction.
    However, \cite{tatman2018practical} suggested that in addition to source code and data availability, clear communication of software dependencies is another important obstacle towards \ac{ml} reproduction on machines that are different from the one where a code was originally developed.
    In addition to code dependencies, we require exact descriptions of algorithmic details.
    To study these aspects, \citet{raff2019step, raff2021research} developed a survival model to predict how long it would take to re-implement and reproduce the results from a paper based on features of the paper, like whether it contains pseudocode, a hyperparameter specification, et cetera.

    Neighboring computational fields have started to identify code quality as a key component to reproducible research~\citep{hoyt2023improving,ziemann2023five,Yasset2016TenGit,Andreas2012TenSoftware,Kjetil2013TenReproducible,Lost2017TenUsable}.
    Similarly, in spirit, this work focuses on best practices for code quality in Section~\ref{sec:dependencies} and Section~\ref{sec:packaging} we discuss best practices like dependency documentation and packaging.

    \paragraph{Operationalization of reproducibility Ideas}\ac{neurips} recently published a reproducibility checklist and updated its code submission guidelines to encourage reproducibility~\citep{pineau2021improving}.
    In our data, we see a small improvement trend around 2020, when the \ac{neurips} code guide~\citep{NeuripsCodeguide} went into effect.
    Section~\ref{subsec:reqs} provides a further elaboration.
    Similarly, \ac{iclr} began asking authors in 2022 \footnote{\url{https://iclr.cc/Conferences/2022/AuthorGuide}, reproducibility in not part of the 2021 guide at \url{https://iclr.cc/Conferences/2021/AuthorGuide}.} to include an optional reproducibility statement, while the \ac{mlrc} encourages investigations into \ac{ml} reproducibility\footnote{\cite{princeton2025reproml}, \url{https://reproml.org/} }.
    In the life sciences, \cite{heil2021reproducibility} introduced tangible criteria in the form of a three-tiered reproducibility scale in which the first level requires data, models, and code to be shared along with the paper.
    The second level requires projects to document software dependencies, the order of commands necessary for reproduction, and to deactivate all stochastic code elements.
    The third level, or \enquote{gold standard}, requires enabling reproduction of a paper's analysis with a single command.

    \paragraph{Criticisms of open source}The \ac{ml} community does not unanimously argue in favor of open source code releases.
    For example, \cite{raff2023siren} expressed concerns that releasing open source code could first lead to a relaxation of standards for detailed descriptions within papers and second could enable divergence between code and paper.
    We suggest that these are instead editorial and peer review issues, which could be alleviated with the improved application of software engineering best practices that both support review and enable automated testing for key assumptions communicated in the paper.

    \paragraph{Reproducibility bolsters citations}Finally, the statistical analysis by \cite{raff2023does} suggested that reproducible articles are cited more frequently.
    .
    Following best practices is, therefore, both in the author's and in the community's best interest.
    The next section describes these best practices.

    \section{Best practices}\label{sec:best_practices}

    We review a subset of software engineering best practices and comment on how they can be applied by the \ac{ml} community.
    Here, we exclusively consider those that can be automatically measured.

    \subsection{Licensing software}\label{sec:licensing}

    A license communicates the terms under which source code can be used, changed, and distributed.
    Without a license, source code can not be (legally) reproduced, modified, nor distributed\footnote{On GitHub, the terms of service apply, minimally allowing all repositories to be viewed and forked, even without a license}.
    A permissive license, such as one suggested by the Open Source Initiative\footnote{\url{https://opensource.org/}}, enables others to improve, reuse, and extend the code~\citep{sonnenburg2007need,Yasset2016TenGit}.
    This can further extend the life of a project since development can continue even without the original authors.
    \cite{sonnenburg2007need} and \url{https://choosealicense.com/} offer guidance for choosing an appropriate license.

    \subsection{Onboarding new users with a README}

    A README file is typically the first documentation that a reader checks in a source code repository.
    It should include a project description, a guide for installation, a quick start guide, and share information on how to contribute~\citep{NeuripsCodeguide,Yasset2016TenGit}.

    Providing a license and README is a language-agnostic step.
    Below, we focus on the specifics of Python, the most common programming language in \ac{ml} research.

    \subsection{Formatting, linting, and type checking}

    Programming language communities often establish style conventions to make source code more uniform and reduce the cognitive burden on readers.
    The Python community suggested best practices in PEP-8~\citep{pep8} and has several tools for automatically formatting code (e.g., Ruff, Black) and for linting code (e.g., Flake8, PyFlakes, Ruff).
    Optional static type hints \citep{vanRossum2014Pep484} enable the implicit documentation of functions as well as the ability to check formal correctness and identify bugs using static type checkers like MyPy \citep{mypy-team2025docs}.

    \subsection{Enumerating dependencies}\label{sec:dependencies}

    Most \ac{ml} projects written in Python depend on external Python code, such as PyTorch.
    Therefore, it is crucial to enumerate these dependencies such that they can be automatically installed.
    A historical approach has been to enumerate the direct dependencies (i.e., those appearing in the \ac{ml} code) in a `requirements.txt' file, which can then be installed via `pip install -r requirements.txt' \citep{pip2025docs,NeuripsCodeguide}.
    The `requirements.txt' has been historically created by running `pip freeze', which outputs a lock file that contains all currently installed Python packages (in the current environment) with version pins following manual installation via `pip install'.

    There have been several iterations of project management tools and configuration formats that attempt to systematize the declaration of direct dependencies, including setuptools, Poetry, Hatch, PDM, and ultimately uv.
    Each tool has historically created its own configuration files (\texttt{setup.py}, \texttt{poetry.toml}, \texttt{hatch.toml}, \texttt{pdm.toml}, \texttt{uv.toml}), which motivated the Python community to define a standard configuration format and filename \texttt{pyproject.toml} in PEP-621~\citep{Cannon2020Pep621}.
    Further, many of these project management tools performed similar locking operations to \texttt{pip freeze}, which created their own lock files (\texttt{Pipfile.lock}, \texttt{poetry.lock}, \texttt{uv.lock}, etc.) that can be used to reproduce an exact environment.
    The Python community defined a standard configuration format and filename \texttt{pylock.toml} in \ac{pep}-751 \citep{Cannon2024Pep751}, which will simplify reproduction in the future.

    Conda~\citep{conda2025repo} is a more generic package manager that can install both Python and non-Python dependencies, then generate a `environment.yml' file that lists dependencies.
    However, Conda is not a packaging tool and is often incorrectly used as a substitute for properly packaging Python code.

    \subsection{Packaging code}\label{sec:packaging}

    Most \ac{ml} projects re-use functions or even entire modules from others.
    While directly copying code is simple, it is problematic for maintainability and visibility~\citep{Yasset2016TenGit}.
    For example, if a bug is identified, it must be manually fixed for every duplication.
    Packaging solves this problem by enabling code to be uploaded to the \ac{pypi}, installed with \ac{pip}, then imported using standard \texttt{import} statements without local code duplication.

    Packaging is relatively straightforward: A packaged project should have \texttt{src} and \texttt{tests} folders \citep{packaging2025python}.
    Metadata about the package, such as its name, license, and dependencies, can be encoded in a standard \texttt{pyproject.toml} file~\citep{Cannon2016pep518,Smith2015pep517}.

    The \texttt{pyproject.toml} also specifies a \textit{build backend} (e.g., \texttt{setuptools}), which produces artifacts for \ac{pypi}~\citep{packaging2025python, Smith2015pep517}.
    When authors use \texttt{setuptools}~\citep{setuptools2025docs} as a build backend, they also need to include a \texttt{setup.py} or \texttt{setup.cfg}.
    Hatch~\citep{hatch2025docs}, Poetry, and uv can be used as a setuptools alternative for users seeking a high-level experience with carefully chosen defaults.

    By packaging our code, we enable others to replicate and build on our work via automatic installation.
    The process gives users access to a working installation now and to package updates and bug fixes in the future.
    Generally, we recommend widespread adoption of the backend-agnostic \texttt{pyproject.toml}-file~\citep{Cannon2016pep518}.

    \subsection{Testing}


    Tests allow us to check new ideas before integrating them into the core codebase~\citep{Yasset2016TenGit}.
    By writing tests, we document core development assumptions and the key behavior of the code.
    When other groups or new people in the same group join a project, the tests help them ensure older features remain operational as new functionality is added.
    Two popular test frameworks for the Python programming language are \texttt{pytest}~\citep{pytest2025docs} and \texttt{unittests}~\citep{unittest2025pythondocs}.
    Regarding test organization, the \ac{pypa} recommends organizing them into a \texttt{tests} folder~\citep{packaging2025python} in the root directory of the project.
    Within the Jax ecosystem, Chex~\citep{chex2025docs} offers an elegant and modular way to automatically evaluate individual tests in different \ac{ml}-specific settings.
    For example, its \texttt{@variants} decorator automatically tests a function's compatibility with \ac{JIT} compilation, device mapping, and parallelization.

    Tests allow maintainers, contributors, and users to verify that the code works as intended, which fosters confidence in new contributors.
    In this setting, successful test runs indicate that everything is set up correctly~\citep{Lost2017TenUsable}.
    In the long run, tests ensure that the code remains functional as it evolves.
    Without it, bugs may be introduced, but not discovered.
    Such hidden errors are a form of technical debt~\citep{breck2017ml}.
    During the development phase, automated tests facilitate code verification and prevent the accumulation of technical debt.
    The next section focuses on the test automation.

    \subsubsection{Automation}

    After configuring formatting, linting, type checking, packaging, and testing for a project, it is possible to standardize, consolidate, and automate their application using workflow tools like \texttt{tox} or \texttt{nox}.
    These tools automate the creation of isolated virtual environments, installation of packaged code, and allow for arbitrary configuration of other build steps that need to be run, e.g., before tests.
    They are configured with a \texttt{tox.toml}/\texttt{tox.ini} or \texttt{noxfile.py}, respectively, which appear in the root directory of the project.

    Finally, \ac{ci} can be used to run these workflow tools, e.g., on all pushes to a project on GitHub.
    On GitHub, this can be configured via a \text{yaml} file placed in the \texttt{.github/workflows} directory.
    Overall, \ac{ci} facilitates enhanced cooperation between team members through automatic verification of new features.

    \subsubsection{Recording seeds}

    Because neural network optimization is typically not a convex problem, we often require pseudorandom initialization to initialize network matrices before we begin experimenting.
    However, in order to be reproducible, stochastic behavior must be disabled and the seed given to the random number generator must be made explicit~\citep{heil2021reproducibility}.
    In this regard, Jax~\citep{jax2018github}, for example, makes the state of the pseudorandom number generator explicit by introducing a unique key object.
    Similarly, PyTorch users can optionally choose to set seed values~\citep{PyTorch2024randomness}.

    \subsection{Documenting code}

    Documentation is the component that makes code accessible to others.
    Typically, documentation is generated in the form of docstrings for every user-facing function, class, and module.
    Automated tools like Sphinx \citep{sphinx2025docs} allow us to generate documentation websites from the docstrings.
    It is customary to create a \texttt{docs} folder for the documentation~\citep{sphinx2025docs}.
    Specialized web services such as ReadTheDocs\footnote{\url{https://about.readthedocs.com/}} automate the build and hosting of documentation.

    \subsection{Difficulties in adopting best practices}\label{sec:difficulties-in-adopting}

    \cite{johanson2018software} observed low adoption rates of modern software engineering techniques in
    computational sciences.
    The work also finds that few scientists are trained in software engineering.
    We suspect a similar situation within the \ac{ml}-community and argue that professors don't place value on it, since they often have no experience working this way.
    Furthermore, working under tight conference deadlines probably exacerbates this problem.
    We find that after a steep initial learning curve, proper software engineering practices are not hard to follow.
    Especially since we can rely on pre-configured templates\footnote{\url{https://cookiecutter.readthedocs.io/en/2.0.2/README.html}}.
    In the long run, especially when considering project handover from one PhD student generation to another, software engineering saves time.

    \section{Methods}
    We developed an automated pipeline that quantifies the adoption of the software engineering best practices described in the previous section.
    Our focus is the \ac{ml} community, where lots of research appears in conference proceedings.
    We generated criteria for choosing \ac{ml} journals and conferences based on their generality and reputation. We limited ourselves to four top conference venues and two journals due to time constraints. The \ac{neurips}, \ac{icml}, \ac{iclr}, and the \ac{aistats} conference, as well as \ac{tmlr} and \ac{mloss}, are included in this study.
    
    The pipeline first downloads \ac{pdf} documents in bulk from select journals and conferences.
    We wrote custom web scrapers when proceedings websites are available.
    Whenever no proceedings had been published yet, we relied on the OpenReview-API to download the proceedings.
    This was the case for ICML 2025 at the time we crawled the data.
    First, for proceedings pages, we use beautiful-soup~\citep{richardson2023soupdocs} to extract all links to papers by filtering for links ending with \texttt{pdf}.
    Second, we extracted links to source code repositories hosted on GitHub.
    We use \texttt{pdfx}~\citep{Hager2021pdfx} and a small number of custom natural language processing functions for \ac{pdf} processing.
    Finally, in each repository, we look for the existence of the following files and their common spelling variants and file extension variants in order to estimate the adoption of software engineering best practices: (\texttt{LICENSE},
    \texttt{COPYING},
    \texttt{README},
    \texttt{requirements.txt},
    \texttt{Pipfile.lock},
    \texttt{pylock.toml},
    \texttt{pyproject.toml},
    \texttt{tox.toml},
    \texttt{tox.ini},
    \texttt{setup.py},
    \texttt{setup.cfg},
    \texttt{noxfile.py},
    \texttt{environment.yml},
    \texttt{uv.lock},
    \texttt{poetry.lock},
    \texttt{poetry.toml},
    \texttt{hatch.toml},
    \texttt{pixi.lock},
    \texttt{pixi.toml},
    \texttt{.pre-commit-config.yaml},
    \texttt{Makefile}) and folders
    (\texttt{docs},
    \texttt{test},
    \texttt{tests},
    \texttt{.github/workflows}).
    Test folders sometimes appear inside the \texttt{src} folder or in a folder with the same name as the project. For completeness, we check these locations as well.

    Before looking at conference proceedings, we first estimate the adoption of software engineering best practices for the journals such as \acf{tmlr} and \acf{mloss} (see Figure~\ref{fig:jmlr_stats}).
    We use \ac{mloss} as a baseline because its submissions are generally reusable software, which in turn correlates with the adoption of software engineering best practices.
    To broaden our view further, we look for signs of best practice adoption within the software repositories we extracted from papers that appeared at major machine learning conferences since 2018.

    Since our analysis is Python-specific, the Python adoption rate should be considered as the upper limit when reading Figure~\ref{fig:jmlr_stats}.
    To avoid confusion, we exclude repositories that do not use Python later in this section, when we discuss Python-specific methods.

    \section{An assessment of the software ecosystem state in \ac{ml} research}\label{sec:estimate_eng_icml}

    \begin{figure}
        \centering
\begin{tikzpicture}

\definecolor{darkgray176}{RGB}{176,176,176}
\definecolor{darkorange25512714}{RGB}{255,127,14}
\definecolor{lightgray204}{RGB}{204,204,204}
\definecolor{steelblue31119180}{RGB}{31,119,180}

\begin{axis}[
legend cell align={left},
legend columns=2,
legend style={fill opacity=0.8, draw opacity=1, text opacity=1, draw=lightgray204},
tick align=outside,
tick pos=left,
title={Estimated adoption},
x grid style={darkgray176},
xmajorgrids,
xmin=-0.45, xmax=6.7,
xtick style={color=black},
xtick={0.25,1.25,2.25,3.25,4.25,5.25,6.25},
xticklabels={README,LICENSE,Python,dependencies,packaged,tests,docs},
x tick label style={rotate=65},
y grid style={darkgray176},
ylabel={Adoption [\%]},
ymajorgrids,
ymin=0, ymax=100,
ytick style={color=black}
]
\draw[draw=none,fill=steelblue31119180] (axis cs:-0.125,0) rectangle (axis cs:0.125,96.1848862802641);
\addlegendimage{ybar,ybar legend,draw=none,fill=steelblue31119180}
\addlegendentry{TMLR}

\draw[draw=none,fill=steelblue31119180] (axis cs:0.875,0) rectangle (axis cs:1.125,50.1834189288335);
\draw[draw=none,fill=steelblue31119180] (axis cs:1.875,0) rectangle (axis cs:2.125,89.6551724137931);
\draw[draw=none,fill=steelblue31119180] (axis cs:2.875,0) rectangle (axis cs:3.125,44.534115920763);
\draw[draw=none,fill=steelblue31119180] (axis cs:3.875,0) rectangle (axis cs:4.125,19.2956713132795);
\draw[draw=none,fill=steelblue31119180] (axis cs:4.875,0) rectangle (axis cs:5.125,7.0432868672047);
\draw[draw=none,fill=steelblue31119180] (axis cs:5.875,0) rectangle (axis cs:6.125,5.64930300807043);
\draw[draw=none,fill=darkorange25512714] (axis cs:0.125,0) rectangle (axis cs:0.375,93.8271604938272);
\addlegendimage{ybar,ybar legend,draw=none,fill=darkorange25512714}
\addlegendentry{MLOSS}

\draw[draw=none,fill=darkorange25512714] (axis cs:1.125,0) rectangle (axis cs:1.375,88.8888888888889);
\draw[draw=none,fill=darkorange25512714] (axis cs:2.125,0) rectangle (axis cs:2.375,80.2469135802469);
\draw[draw=none,fill=darkorange25512714] (axis cs:3.125,0) rectangle (axis cs:3.375,29.6296296296296);
\draw[draw=none,fill=darkorange25512714] (axis cs:4.125,0) rectangle (axis cs:4.375,76.5432098765432);
\draw[draw=none,fill=darkorange25512714] (axis cs:5.125,0) rectangle (axis cs:5.375,67.2839506172839);
\draw[draw=none,fill=darkorange25512714] (axis cs:6.125,0) rectangle (axis cs:6.375,80.2469135802469);
\end{axis}

\end{tikzpicture}
        \caption{Estimated state of software engineering best practices at \ac{tmlr} and \ac{mloss}. The software focused \acs{mloss} serves as a baseline for comparison. The plots illustrate web-crawled percentages of files and folders tied to the adoption of software engineering best practices.}
        \label{fig:jmlr_stats}
    \end{figure}

    In Figure~\ref{fig:jmlr_stats}, we observe that \ac{tmlr} papers are not yet at the level of \ac{mloss} in terms of best practices adoption.
    We observe growth potential for \ac{tmlr} code submissions in almost all dimensions in comparison to \ac{mloss}.
    The rest of this section considers conference venues where most \ac{ml}
    research appears, we will keep the \ac{mloss}-baseline in mind.

    \paragraph{README and LICENSE files over time}

    \begin{figure}
        \centering
\begin{tikzpicture}

\definecolor{crimson2143940}{RGB}{214,39,40}
\definecolor{darkgray176}{RGB}{176,176,176}
\definecolor{darkorange25512714}{RGB}{255,127,14}
\definecolor{forestgreen4416044}{RGB}{44,160,44}
\definecolor{lightgray204}{RGB}{204,204,204}
\definecolor{steelblue31119180}{RGB}{31,119,180}

\begin{axis}[
legend cell align={left},
legend style={
  fill opacity=0.8,
  draw opacity=1,
  text opacity=1,
  at={(0.97,0.03)},
  anchor=south east,
  draw=lightgray204
},
tick align=outside,
tick pos=left,
title={README},
x grid style={darkgray176},
xlabel={conference year},
xmajorgrids,
xmin=0.6, xmax=8.4,
xtick style={color=black},
xtick={1,2,3,4,5,6,7,8},
xticklabels={18,19,20,21,22,23,24,25},
y grid style={darkgray176},
ylabel={adoption [\%]},
ymajorgrids,
ymin=91.340206185567, ymax=100.412371134021,
ytick style={color=black}
]
\addplot [semithick, steelblue31119180, mark=*, mark size=3, mark options={solid}]
table {%
1 96.875
2 96.9298245614035
3 96.9616908850727
4 98.8112927191679
5 98.5024958402662
6 98.0728051391863
7 98.8239247311828
8 98.5664985399522
};
\addlegendentry{icml}
\addplot [semithick, darkorange25512714, mark=*, mark size=3, mark options={solid}]
table {%
1 93.75
2 95.2941176470588
3 95.1807228915663
4 94.4915254237288
5 96.085409252669
6 96.8115942028986
7 94.3310657596372
8 96.1693548387097
};
\addlegendentry{aistats}
\addplot [semithick, forestgreen4416044, mark=*, mark size=3, mark options={solid}]
table {%
1 94.6428571428571
2 97.196261682243
3 98.6842105263158
4 98.9045383411581
5 98.7442922374429
6 99.044151115157
7 98.46875
8 98.8959907030796
};
\addlegendentry{iclr}
\addplot [semithick, crimson2143940, mark=*, mark size=3, mark options={solid}]
table {%
1 96.6494845360825
2 97.0282317979198
3 98.1302774427021
4 99.1499227202473
5 98.3952171176841
6 98.7316287497483
7 98.6107005616317
};
\addlegendentry{neurips}
\end{axis}

\end{tikzpicture}
\begin{tikzpicture}

\definecolor{crimson2143940}{RGB}{214,39,40}
\definecolor{darkgray176}{RGB}{176,176,176}
\definecolor{darkorange25512714}{RGB}{255,127,14}
\definecolor{forestgreen4416044}{RGB}{44,160,44}
\definecolor{lightgray204}{RGB}{204,204,204}
\definecolor{steelblue31119180}{RGB}{31,119,180}

\begin{axis}[
legend cell align={left},
legend style={
  fill opacity=0.8,
  draw opacity=1,
  text opacity=1,
  at={(0.97,0.03)},
  anchor=south east,
  draw=lightgray204
},
tick align=outside,
tick pos=left,
title={LICENSE},
x grid style={darkgray176},
xlabel={conference year},
xmajorgrids,
xmin=0.6, xmax=8.4,
xtick style={color=black},
xtick={1,2,3,4,5,6,7,8},
xticklabels={18,19,20,21,22,23,24,25},
y grid style={darkgray176},
ylabel={adoption [\%]},
ymajorgrids,
ymin=40.7258771929825, ymax=81.2828947368421,
ytick style={color=black}
]
\addplot [semithick, steelblue31119180, mark=*, mark size=3, mark options={solid}]
table {%
1 59.765625
2 64.9122807017544
3 65.5217965653897
4 65.8246656760773
5 68.2196339434276
6 65.1498929336188
7 64.3145161290323
8 65.1446774621715
};
\addlegendentry{icml}
\addplot [semithick, darkorange25512714, mark=*, mark size=3, mark options={solid}]
table {%
1 56.25
2 56.4705882352941
3 50
4 52.9661016949153
5 60.4982206405694
6 52.7536231884058
7 51.4739229024943
8 54.2338709677419
};
\addlegendentry{aistats}
\addplot [semithick, forestgreen4416044, mark=*, mark size=3, mark options={solid}]
table {%
1 58.9285714285714
2 75.7009345794392
3 73.4210526315789
4 73.8654147104851
5 72.4885844748858
6 71.279016841147
7 74.375
8 68.4485764090645
};
\addlegendentry{iclr}
\addplot [semithick, crimson2143940, mark=*, mark size=3, mark options={solid}]
table {%
1 62.1134020618557
2 57.8008915304606
3 64.2340168878167
4 68.7017001545595
5 67.1176840780365
6 68.1095228508154
7 67.1741058232338
};
\addlegendentry{neurips}
\end{axis}

\end{tikzpicture}
        \caption{Estimated state of README and LICENSE file adoption in major \ac{ml} conferences. We add the counts for \texttt{README.md} and \texttt{README.rst} files, as well as common spelling variations and show these as \texttt{README}. For the licenses, we add the counts for \texttt{LICENSE}, \texttt{COPYING}, and common spelling variations.}
        \label{fig:license_and_readme}
    \end{figure}

    In Figure~\ref{fig:license_and_readme}, we observe that the inclusion of README files is widespread, with nearly all repositories across all journals and conferences having full adoption.
    However, the inclusion of license files seems to have stagnated between 50\% and 80\% over time, with \ac{iclr}, \ac{neurips}, and \ac{icml} having the most.
    This means users are potentially working without legal security with the implications outlined in section~\ref{sec:licensing}.

    \paragraph{Python adoption over time}

    \begin{figure}
        \centering
\begin{tikzpicture}

\definecolor{crimson2143940}{RGB}{214,39,40}
\definecolor{darkgray176}{RGB}{176,176,176}
\definecolor{darkorange25512714}{RGB}{255,127,14}
\definecolor{forestgreen4416044}{RGB}{44,160,44}
\definecolor{lightgray204}{RGB}{204,204,204}
\definecolor{steelblue31119180}{RGB}{31,119,180}

\begin{axis}[
legend cell align={left},
legend style={
  fill opacity=0.8,
  draw opacity=1,
  text opacity=1,
  at={(0.97,0.03)},
  anchor=south east,
  draw=lightgray204
},
tick align=outside,
tick pos=left,
title={Python use},
x grid style={darkgray176},
xlabel={conference year},
xmajorgrids,
xmin=0.6, xmax=8.4,
xtick style={color=black},
xtick={1,2,3,4,5,6,7,8},
xticklabels={18,19,20,21,22,23,24,25},
y grid style={darkgray176},
ylabel={adoption [\%]},
ymajorgrids,
ymin=60.9638908916126, ymax=97.2846070656092,
ytick style={color=black}
]
\addplot [semithick, steelblue31119180, mark=*, mark size=3, mark options={solid}]
table {%
1 82.03125
2 85.0877192982456
3 91.1492734478203
4 91.8276374442793
5 92.5956738768719
6 92.0235546038544
7 93.6155913978495
8 87.6294133262543
};
\addlegendentry{icml}
\addplot [semithick, darkorange25512714, mark=*, mark size=3, mark options={solid}]
table {%
1 70.8333333333333
2 80
3 78.9156626506024
4 86.864406779661
5 84.6975088967972
6 86.0869565217391
7 83.9002267573696
8 88.3064516129032
};
\addlegendentry{aistats}
\addplot [semithick, forestgreen4416044, mark=*, mark size=3, mark options={solid}]
table {%
1 83.9285714285714
2 91.1214953271028
3 93.6842105263158
4 92.8012519561815
5 94.4063926940639
6 93.5821574874829
7 93.40625
8 90.9355026147589
};
\addlegendentry{iclr}
\addplot [semithick, crimson2143940, mark=*, mark size=3, mark options={solid}]
table {%
1 85.3092783505155
2 85.5869242199108
3 90.2291917973462
4 93.2380216383308
5 91.5984896161108
6 91.9065834507751
7 92.0780372450488
};
\addlegendentry{neurips}
\end{axis}

\end{tikzpicture}
\begin{tikzpicture}

\definecolor{crimson2143940}{RGB}{214,39,40}
\definecolor{darkgray176}{RGB}{176,176,176}
\definecolor{darkorange25512714}{RGB}{255,127,14}
\definecolor{forestgreen4416044}{RGB}{44,160,44}
\definecolor{lightgray204}{RGB}{204,204,204}
\definecolor{steelblue31119180}{RGB}{31,119,180}

\begin{axis}[
legend cell align={left},
legend style={
  fill opacity=0.8,
  draw opacity=1,
  text opacity=1,
  at={(0.97,0.03)},
  anchor=south east,
  draw=lightgray204
},
tick align=outside,
tick pos=left,
title={docs folder},
x grid style={darkgray176},
xlabel={conference year},
xmajorgrids,
xmin=0.6, xmax=8.4,
xtick style={color=black},
xtick={1,2,3,4,5,6,7,8},
xticklabels={18,19,20,21,22,23,24,25},
y grid style={darkgray176},
ylabel={adoption [\%]},
ymajorgrids,
ymin=6.44232055605807, ymax=23.4741549207186,
ytick style={color=black}
]
\addplot [semithick, steelblue31119180, mark=*, mark size=3, mark options={solid}]
table {%
1 10.546875
2 8.7719298245614
3 15.4557463672391
4 17.3848439821694
5 18.8851913477537
6 16.541755888651
7 17.4395161290323
8 19.0868064773029
};
\addlegendentry{icml}
\addplot [semithick, darkorange25512714, mark=*, mark size=3, mark options={solid}]
table {%
1 12.5
2 17.6470588235294
3 10.2409638554217
4 15.2542372881356
5 16.7259786476868
6 12.463768115942
7 12.9251700680272
8 12.9032258064516
};
\addlegendentry{aistats}
\addplot [semithick, forestgreen4416044, mark=*, mark size=3, mark options={solid}]
table {%
1 10.2678571428571
2 12.1495327102804
3 16.5789473684211
4 16.1189358372457
5 15.5251141552511
6 21.8934911242604
7 21.28125
8 22.6999806314158
};
\addlegendentry{iclr}
\addplot [semithick, crimson2143940, mark=*, mark size=3, mark options={solid}]
table {%
1 10.0515463917526
2 13.9673105497771
3 15.9831121833534
4 18.3153013910356
5 19.3517935808685
6 20.032212603181
7 19.1693762932309
};
\addlegendentry{neurips}
\end{axis}

\end{tikzpicture}
        \caption{Repository link crawl results for Python adoption (left) and standalone documentation (right).
        We count each repository where Python is listed as a language in GitHub's language box. The plot on the right of this figure illustrates the share of repositories with a \texttt{doc} or \texttt{docs} folder.}
        \label{fig:uses_python_docs}
    \end{figure}

    Most of the engineering best practices we described previously in Section~\ref{sec:best_practices} are Python specific, before moving on, we must therefore check if Python is indeed the most common language in \ac{ml}.
    In the left-hand plot of Figure~\ref{fig:uses_python_docs}, we observe an upwards trend in the usage of Python over time.
    Since 2021, more than 80\% of repositories use Python at all major conferences we considered.
    We believe this trend bolsters the case for more rigorous software engineering.
    Since fewer language barriers exist, packaging especially would help us to collaborate more effectively as a community.
    After all, since almost everyone is working with Python, the \texttt{import} statement is available to the vast majority of the community, and can help us to avoid code duplication.

    The following sections will focus on Python-specific engineering practices.
    We exclude repositories in other languages from our analysis.

    \paragraph{Documentation}

    In the right-hand side of Figure~\ref{fig:uses_python_docs}, we observe that the share of projects with a standalone documentation folder approaches 20\%.
    While this share remains low, we see a clear upward trend, and in some cases, a well-written README file is also sufficient.

    \paragraph{Linting}
    \begin{figure}
        \centering
\begin{tikzpicture}

\definecolor{crimson2143940}{RGB}{214,39,40}
\definecolor{darkgray176}{RGB}{176,176,176}
\definecolor{darkorange25512714}{RGB}{255,127,14}
\definecolor{forestgreen4416044}{RGB}{44,160,44}
\definecolor{lightgray204}{RGB}{204,204,204}
\definecolor{steelblue31119180}{RGB}{31,119,180}

\begin{axis}[
legend cell align={left},
legend style={
  fill opacity=0.8,
  draw opacity=1,
  text opacity=1,
  at={(0.03,0.97)},
  anchor=north west,
  draw=lightgray204
},
tick align=outside,
tick pos=left,
title={.flake8},
x grid style={darkgray176},
xlabel={conference year},
xmajorgrids,
xmin=-0.35, xmax=7.35,
xtick style={color=black},
xtick={0,1,2,3,4,5,6,7},
xticklabels={18,19,20,21,22,23,24,25},
y grid style={darkgray176},
ylabel={adoption [\%]},
ymajorgrids,
ymin=-0.169286577992745, ymax=3.55501813784764,
ytick style={color=black}
]
\addplot [semithick, steelblue31119180, mark=*, mark size=3, mark options={solid}]
table {%
0 0
1 1.28865979381443
2 2.02898550724638
3 1.45631067961165
4 2.42587601078167
5 1.97789412449098
6 1.90236898779612
7 2.42350802787034
};
\addlegendentry{icml}
\addplot [semithick, darkorange25512714, mark=*, mark size=3, mark options={solid}]
table {%
0 0
1 0
2 0.763358778625954
3 0.48780487804878
4 0.840336134453782
5 1.68350168350168
6 0.810810810810811
7 2.05479452054794
};
\addlegendentry{aistats}
\addplot [semithick, forestgreen4416044, mark=*, mark size=3, mark options={solid}]
table {%
0 0
1 0
2 0
3 2.24719101123596
4 1.18043844856661
5 3.3857315598549
6 3.16147859922179
7 3.1197015937606
};
\addlegendentry{iclr}
\addplot [semithick, crimson2143940, mark=*, mark size=3, mark options={solid}]
table {%
0 0.604229607250755
1 1.90972222222222
2 2.40641711229947
3 2.73518441773726
4 2.9886636894538
5 2.71631982475356
6 2.48796147672552
};
\addlegendentry{neurips}
\end{axis}

\end{tikzpicture}
        \caption{Adoption of \texttt{.flake8} configuration files in repositories over time.}
        \label{fig:falke}
    \end{figure}

    In Figure~\ref{fig:falke}, we observe a relatively low, but upwards-trending, adoption of configuration for Flake8.
    However, tracking \texttt{.flake8} is not necessarily representative of linting adoption because of widespread adoption of Ruff\footnote{\url{https://github.com/astral-sh/ruff}}, which is typically configured in the \texttt{pyproject.toml}.
    Therefore, we can estimate an upper bound on linting by combining \texttt{pyproject.toml} and \texttt{setup.cfg} file counts from Figure~\ref{fig:packaging} with the \texttt{.flake8} numbers from Figure~\ref{fig:falke}, which places us well under 50\%.

    \paragraph{Requirements documentation}\label{subsec:reqs}

    \begin{figure}
        \centering
\begin{tikzpicture}

\definecolor{crimson2143940}{RGB}{214,39,40}
\definecolor{darkgray176}{RGB}{176,176,176}
\definecolor{darkorange25512714}{RGB}{255,127,14}
\definecolor{forestgreen4416044}{RGB}{44,160,44}
\definecolor{lightgray204}{RGB}{204,204,204}
\definecolor{steelblue31119180}{RGB}{31,119,180}

\begin{axis}[
legend cell align={left},
legend style={
  fill opacity=0.8,
  draw opacity=1,
  text opacity=1,
  at={(0.03,0.97)},
  anchor=north west,
  draw=lightgray204
},
tick align=outside,
tick pos=left,
title={requirements.txt},
x grid style={darkgray176},
xlabel={conference year},
xmajorgrids,
xmin=-0.35, xmax=7.35,
xtick style={color=black},
xtick={0,1,2,3,4,5,6,7},
xticklabels={18,19,20,21,22,23,24,25},
y grid style={darkgray176},
ylabel={adoption [\%]},
ymajorgrids,
ymin=7.15721181656793, ymax=43.816198910897,
ytick style={color=black}
]
\addplot [semithick, steelblue31119180, mark=*, mark size=3, mark options={solid}]
table {%
0 18.5714285714286
1 17.5257731958763
2 29.1304347826087
3 31.8770226537217
4 31.1769991015274
5 34.9621873182083
6 40.2727925340991
7 40.9875795213572
};
\addlegendentry{icml}
\addplot [semithick, darkorange25512714, mark=*, mark size=3, mark options={solid}]
table {%
0 8.82352941176471
1 22.0588235294118
2 25.9541984732824
3 21.9512195121951
4 24.7899159663866
5 30.3030303030303
6 32.7027027027027
7 36.5296803652968
};
\addlegendentry{aistats}
\addplot [semithick, forestgreen4416044, mark=*, mark size=3, mark options={solid}]
table {%
0 18.75
1 19.1489361702128
2 26.6666666666667
3 24.438202247191
4 26.6441821247892
5 31.1970979443773
6 32.2470817120623
7 42.1498813157002
};
\addlegendentry{iclr}
\addplot [semithick, crimson2143940, mark=*, mark size=3, mark options={solid}]
table {%
0 19.9395770392749
1 27.6041666666667
2 29.144385026738
3 34.1898052217157
4 35.4860872552388
5 37.2179627601314
6 42.1348314606742
};
\addlegendentry{neurips}
\end{axis}

\end{tikzpicture}
\begin{tikzpicture}

\definecolor{crimson2143940}{RGB}{214,39,40}
\definecolor{darkgray176}{RGB}{176,176,176}
\definecolor{darkorange25512714}{RGB}{255,127,14}
\definecolor{forestgreen4416044}{RGB}{44,160,44}
\definecolor{lightgray204}{RGB}{204,204,204}
\definecolor{steelblue31119180}{RGB}{31,119,180}

\begin{axis}[
legend cell align={left},
legend style={
  fill opacity=0.8,
  draw opacity=1,
  text opacity=1,
  at={(0.03,0.97)},
  anchor=north west,
  draw=lightgray204
},
tick align=outside,
tick pos=left,
title={environment},
x grid style={darkgray176},
xlabel={conference year},
xmajorgrids,
xmin=-0.35, xmax=7.35,
xtick style={color=black},
xtick={0,1,2,3,4,5,6,7},
xticklabels={18,19,20,21,22,23,24,25},
y grid style={darkgray176},
ylabel={adoption [\%]},
ymajorgrids,
ymin=-0.488215488215488, ymax=10.2525252525253,
ytick style={color=black}
]
\addplot [semithick, steelblue31119180, mark=*, mark size=3, mark options={solid}]
table {%
0 3.80952380952381
1 2.06185567010309
2 3.6231884057971
3 5.8252427184466
4 4.94159928122192
5 6.45724258289703
6 7.2505384063173
7 7.30081793395941
};
\addlegendentry{icml}
\addplot [semithick, darkorange25512714, mark=*, mark size=3, mark options={solid}]
table {%
0 5.88235294117647
1 2.94117647058824
2 4.58015267175572
3 2.4390243902439
4 2.52100840336134
5 9.76430976430976
6 5.94594594594595
7 6.62100456621005
};
\addlegendentry{aistats}
\addplot [semithick, forestgreen4416044, mark=*, mark size=3, mark options={solid}]
table {%
0 0
1 1.06382978723404
2 1.02564102564103
3 4.49438202247191
4 2.52951096121417
5 6.16686819830713
6 5.3988326848249
7 8.03662258392675
};
\addlegendentry{iclr}
\addplot [semithick, crimson2143940, mark=*, mark size=3, mark options={solid}]
table {%
0 2.41691842900302
1 3.81944444444444
2 4.81283422459893
3 5.51181102362205
4 6.52696667811749
5 8.30230010952902
6 7.9454253611557
};
\addlegendentry{neurips}
\end{axis}

\end{tikzpicture}
        \caption{Requirements documentation over time. The figure illustrates the share of repositories with \texttt{requirements.txt} and \texttt{environment.yml} files.}    \label{fig:requirements_and_env}
\begin{tikzpicture}

\definecolor{crimson2143940}{RGB}{214,39,40}
\definecolor{darkgray176}{RGB}{176,176,176}
\definecolor{darkorange25512714}{RGB}{255,127,14}
\definecolor{forestgreen4416044}{RGB}{44,160,44}
\definecolor{lightgray204}{RGB}{204,204,204}
\definecolor{steelblue31119180}{RGB}{31,119,180}

\begin{axis}[
legend cell align={left},
legend style={
  fill opacity=0.8,
  draw opacity=1,
  text opacity=1,
  at={(0.03,0.97)},
  anchor=north west,
  draw=lightgray204
},
tick align=outside,
tick pos=left,
title={poetry.lock},
x grid style={darkgray176},
xlabel={conference year},
xmajorgrids,
xmin=-0.35, xmax=7.35,
xtick style={color=black},
xtick={0,1,2,3,4,5,6,7},
xticklabels={18,19,20,21,22,23,24,25},
y grid style={darkgray176},
ylabel={adoption [\%]},
ymajorgrids,
ymin=-0.135135135135135, ymax=2.83783783783784,
ytick style={color=black}
]
\addplot [semithick, steelblue31119180, mark=*, mark size=3, mark options={solid}]
table {%
0 0
1 0.515463917525773
2 0.289855072463768
3 0.647249190938511
4 0.539083557951483
5 0.756253635834788
6 0.933237616654702
7 1.45410481672221
};
\addlegendentry{icml}
\addplot [semithick, darkorange25512714, mark=*, mark size=3, mark options={solid}]
table {%
0 0
1 0
2 0
3 0
4 0
5 1.01010101010101
6 2.7027027027027
7 1.59817351598174
};
\addlegendentry{aistats}
\addplot [semithick, forestgreen4416044, mark=*, mark size=3, mark options={solid}]
table {%
0 0
1 0
2 0
3 0
4 0
5 0.120918984280532
6 0.680933852140078
7 1.32248219735504
};
\addlegendentry{iclr}
\addplot [semithick, crimson2143940, mark=*, mark size=3, mark options={solid}]
table {%
0 0.302114803625378
1 0
2 0.334224598930481
3 0.49730625777041
4 0.893163861216077
5 0.788608981380066
6 1.23595505617978
};
\addlegendentry{neurips}
\end{axis}

\end{tikzpicture}
\begin{tikzpicture}

\definecolor{crimson2143940}{RGB}{214,39,40}
\definecolor{darkgray176}{RGB}{176,176,176}
\definecolor{darkorange25512714}{RGB}{255,127,14}
\definecolor{forestgreen4416044}{RGB}{44,160,44}
\definecolor{lightgray204}{RGB}{204,204,204}
\definecolor{steelblue31119180}{RGB}{31,119,180}

\begin{axis}[
legend cell align={left},
legend style={
  fill opacity=0.8,
  draw opacity=1,
  text opacity=1,
  at={(0.03,0.97)},
  anchor=north west,
  draw=lightgray204
},
tick align=outside,
tick pos=left,
title={uv.lock},
x grid style={darkgray176},
xlabel={conference year},
xmajorgrids,
xmin=-0.35, xmax=7.35,
xtick style={color=black},
xtick={0,1,2,3,4,5,6,7},
xticklabels={18,19,20,21,22,23,24,25},
y grid style={darkgray176},
ylabel={adoption [\%]},
ymajorgrids,
ymin=-0.084822780975462, ymax=1.7812784004847,
ytick style={color=black}
]
\addplot [semithick, steelblue31119180, mark=*, mark size=3, mark options={solid}]
table {%
0 0
1 0
2 0
3 0.485436893203883
4 0.628930817610063
5 0.58173356602676
6 1.04091888011486
7 1.69645561950924
};
\addlegendentry{icml}
\addplot [semithick, darkorange25512714, mark=*, mark size=3, mark options={solid}]
table {%
0 0
1 0
2 0
3 0.48780487804878
4 0
5 0
6 1.08108108108108
7 1.14155251141552
};
\addlegendentry{aistats}
\addplot [semithick, forestgreen4416044, mark=*, mark size=3, mark options={solid}]
table {%
0 0
1 0
2 0
3 0
4 0
5 0
6 0.389105058365759
7 1.18684299762631
};
\addlegendentry{iclr}
\addplot [semithick, crimson2143940, mark=*, mark size=3, mark options={solid}]
table {%
0 0
1 0.173611111111111
2 0.0668449197860962
3 0.331537505180274
4 0.515286843009275
5 0.766703176341731
6 0.995184590690209
};
\addlegendentry{neurips}
\end{axis}

\end{tikzpicture}
        \caption{Overview of the lock files the crawler discovered. The environment plot adds the numbers for
        \texttt{environment.yml} as well as \texttt{environment.yaml} files.}
        \label{fig:lock_files}
    \end{figure}

    Requirements are very important for reproducibility since initializations, for example, differ
    between PyTorch versions\footnote{\url{https://docs.pytorch.org/docs/stable/notes/randomness.html}}.
    The exact version used in a project should therefore appear in the requirements documentation.
    In Figure~\ref{fig:requirements_and_env} and Figure~\ref{fig:lock_files} depict the numbers.
    We observe moderate, upwards-trending adoption of various mechanisms for declaring dependencies.

    The number of \texttt{.lock}-files we found was very small, the \texttt{uv} version appeared most (see supplementary Figure~\ref{fig:lock_other} for the rare ones).
    Lockfiles are higher-level application-centric alternatives to the \texttt{requirements.txt} file.

    Both \texttt{requirements.txt} and environment files appear more frequently.
    At \ac{icml} in 2024, the combined share of projects with \texttt{requirements.txt} and \texttt{environment.yml} files was still less than 50\%.
    This adoption rate implies that the results of many projects will not be reproducible straightforwardly.
    The \ac{neurips} code guide appeared~\citep{NeuripsCodeguide} in 2020, it asks authors to provide these files.
    We see a solid positive trend since the guide appeared. Ideally, we should aim to allow straightforward reproduction for every project.

    It is possible to provide both files automatically with a single command each.
    Pip users can run \texttt{pip freeze > requirements.txt} to create the historic file.
    For the modern standard \texttt{pip lock -e .} will create the \texttt{pylock.toml} file~\citep{Cannon2024Pep751}.
    While we did not find many usages \texttt{pylock.toml} due to its recent introduction, we expect to see more of these files in the future.
    Similarly, for Conda users, providing the file requires typing \texttt{conda env export > environment.yml} into the terminal.
    Afterward, both groups could commit the files to their code repositories, doing so will ensure future scientists end up with the correct software versions.

    \paragraph{Packaging adoption}

    \begin{figure}
        \centering
\begin{tikzpicture}

\definecolor{crimson2143940}{RGB}{214,39,40}
\definecolor{darkgray176}{RGB}{176,176,176}
\definecolor{darkorange25512714}{RGB}{255,127,14}
\definecolor{forestgreen4416044}{RGB}{44,160,44}
\definecolor{lightgray204}{RGB}{204,204,204}
\definecolor{steelblue31119180}{RGB}{31,119,180}

\begin{axis}[
legend cell align={left},
legend style={fill opacity=0.8, draw opacity=1, text opacity=1, draw=lightgray204},
tick align=outside,
tick pos=left,
title={setup.py},
x grid style={darkgray176},
xlabel={conference year},
xmajorgrids,
xmin=-0.35, xmax=7.35,
xtick style={color=black},
xtick={0,1,2,3,4,5,6,7},
xticklabels={18,19,20,21,22,23,24,25},
y grid style={darkgray176},
ylabel={adoption [\%]},
ymajorgrids,
ymin=9.52284105131414, ymax=45.7650187734668,
ytick style={color=black}
]
\addplot [semithick, steelblue31119180, mark=*, mark size=3, mark options={solid}]
table {%
0 31.9047619047619
1 26.8041237113402
2 31.7391304347826
3 32.0388349514563
4 27.9424977538185
5 28.0977312390925
6 26.6690595836324
7 27.3250530142381
};
\addlegendentry{icml}
\addplot [semithick, darkorange25512714, mark=*, mark size=3, mark options={solid}]
table {%
0 44.1176470588235
1 29.4117647058824
2 29.7709923664122
3 33.1707317073171
4 28.9915966386555
5 22.8956228956229
6 27.5675675675676
7 21.4611872146119
};
\addlegendentry{aistats}
\addplot [semithick, forestgreen4416044, mark=*, mark size=3, mark options={solid}]
table {%
0 25
1 11.1702127659574
2 20
3 27.247191011236
4 27.9932546374368
5 29.6251511487303
6 34.2898832684825
7 30.0101729399797
};
\addlegendentry{iclr}
\addplot [semithick, crimson2143940, mark=*, mark size=3, mark options={solid}]
table {%
0 22.0543806646526
1 23.0902777777778
2 26.7379679144385
3 27.8077082469954
4 27.997251803504
5 28.9156626506024
6 26.1797752808989
};
\addlegendentry{neurips}
\end{axis}

\end{tikzpicture}
\begin{tikzpicture}

\definecolor{crimson2143940}{RGB}{214,39,40}
\definecolor{darkgray176}{RGB}{176,176,176}
\definecolor{darkorange25512714}{RGB}{255,127,14}
\definecolor{forestgreen4416044}{RGB}{44,160,44}
\definecolor{lightgray204}{RGB}{204,204,204}
\definecolor{steelblue31119180}{RGB}{31,119,180}

\begin{axis}[
legend cell align={left},
legend style={
  fill opacity=0.8,
  draw opacity=1,
  text opacity=1,
  at={(0.03,0.97)},
  anchor=north west,
  draw=lightgray204
},
tick align=outside,
tick pos=left,
title={pyproject.toml},
x grid style={darkgray176},
xlabel={conference year},
xmajorgrids,
xmin=-0.35, xmax=7.35,
xtick style={color=black},
xtick={0,1,2,3,4,5,6,7},
xticklabels={18,19,20,21,22,23,24,25},
y grid style={darkgray176},
ylabel={adoption [\%]},
ymajorgrids,
ymin=-1.2056952438655, ymax=25.3196001211754,
ytick style={color=black}
]
\addplot [semithick, steelblue31119180, mark=*, mark size=3, mark options={solid}]
table {%
0 3.80952380952381
1 4.38144329896907
2 7.53623188405797
3 13.1067961165049
4 11.9496855345912
5 13.0890052356021
6 16.5829145728643
7 24.1139048773099
};
\addlegendentry{icml}
\addplot [semithick, darkorange25512714, mark=*, mark size=3, mark options={solid}]
table {%
0 2.94117647058824
1 2.94117647058824
2 3.05343511450382
3 11.219512195122
4 10.9243697478992
5 13.8047138047138
6 13.7837837837838
7 16.2100456621005
};
\addlegendentry{aistats}
\addplot [semithick, forestgreen4416044, mark=*, mark size=3, mark options={solid}]
table {%
0 0
1 6.91489361702128
2 4.61538461538462
3 6.17977528089888
4 9.106239460371
5 8.94800483675937
6 14.6887159533074
7 17.9043743641912
};
\addlegendentry{iclr}
\addplot [semithick, crimson2143940, mark=*, mark size=3, mark options={solid}]
table {%
0 2.41691842900302
1 5.38194444444444
2 7.28609625668449
3 9.94612515540821
4 11.7485400206115
5 15.5093099671413
6 18.5072231139647
};
\addlegendentry{neurips}
\end{axis}

\end{tikzpicture}
\begin{tikzpicture}

\definecolor{crimson2143940}{RGB}{214,39,40}
\definecolor{darkgray176}{RGB}{176,176,176}
\definecolor{darkorange25512714}{RGB}{255,127,14}
\definecolor{forestgreen4416044}{RGB}{44,160,44}
\definecolor{lightgray204}{RGB}{204,204,204}
\definecolor{steelblue31119180}{RGB}{31,119,180}

\begin{axis}[
legend cell align={left},
legend style={fill opacity=0.8, draw opacity=1, text opacity=1, draw=lightgray204},
tick align=outside,
tick pos=left,
title={setup.cfg},
x grid style={darkgray176},
xlabel={conference year},
xmajorgrids,
xmin=-0.35, xmax=7.35,
xtick style={color=black},
xtick={0,1,2,3,4,5,6,7},
xticklabels={18,19,20,21,22,23,24,25},
y grid style={darkgray176},
ylabel={adoption [\%]},
ymajorgrids,
ymin=3.73279098873592, ymax=15.2284105131414,
ytick style={color=black}
]
\addplot [semithick, steelblue31119180, mark=*, mark size=3, mark options={solid}]
table {%
0 12.3809523809524
1 8.76288659793814
2 8.98550724637681
3 6.63430420711974
4 7.45732255166217
5 8.20244328097731
6 6.46087580760948
7 6.8767040290821
};
\addlegendentry{icml}
\addplot [semithick, darkorange25512714, mark=*, mark size=3, mark options={solid}]
table {%
0 14.7058823529412
1 11.7647058823529
2 9.9236641221374
3 6.34146341463415
4 6.72268907563025
5 8.41750841750842
6 5.94594594594595
7 6.84931506849315
};
\addlegendentry{aistats}
\addplot [semithick, forestgreen4416044, mark=*, mark size=3, mark options={solid}]
table {%
0 6.25
1 4.25531914893617
2 8.2051282051282
3 9.26966292134832
4 7.92580101180438
5 5.80411124546554
6 9.82490272373541
7 7.05323838589352
};
\addlegendentry{iclr}
\addplot [semithick, crimson2143940, mark=*, mark size=3, mark options={solid}]
table {%
0 6.04229607250755
1 7.29166666666667
2 6.81818181818182
3 8.74430169912971
4 8.58811405015459
5 8.039430449069
6 6.62921348314607
};
\addlegendentry{neurips}
\end{axis}

\end{tikzpicture}
\begin{tikzpicture}

\definecolor{crimson2143940}{RGB}{214,39,40}
\definecolor{darkgray176}{RGB}{176,176,176}
\definecolor{darkorange25512714}{RGB}{255,127,14}
\definecolor{forestgreen4416044}{RGB}{44,160,44}
\definecolor{lightgray204}{RGB}{204,204,204}
\definecolor{steelblue31119180}{RGB}{31,119,180}

\begin{axis}[
legend cell align={left},
legend style={fill opacity=0.8, draw opacity=1, text opacity=1, draw=lightgray204},
tick align=outside,
tick pos=left,
title={src},
x grid style={darkgray176},
xlabel={conference year},
xmajorgrids,
xmin=-0.35, xmax=7.35,
xtick style={color=black},
xtick={0,1,2,3,4,5,6,7},
xticklabels={18,19,20,21,22,23,24,25},
y grid style={darkgray176},
ylabel={adoption [\%]},
ymajorgrids,
ymin=0.606617647058823, ymax=19.6139705882353,
ytick style={color=black}
]
\addplot [semithick, steelblue31119180, mark=*, mark size=3, mark options={solid}]
table {%
0 5.71428571428571
1 6.18556701030928
2 6.52173913043478
3 6.63430420711974
4 7.63701707097933
5 9.94764397905759
6 9.08111988513999
7 9.39109360799758
};
\addlegendentry{icml}
\addplot [semithick, darkorange25512714, mark=*, mark size=3, mark options={solid}]
table {%
0 8.82352941176471
1 1.47058823529412
2 12.9770992366412
3 5.36585365853659
4 10.0840336134454
5 10.4377104377104
6 11.8918918918919
7 14.3835616438356
};
\addlegendentry{aistats}
\addplot [semithick, forestgreen4416044, mark=*, mark size=3, mark options={solid}]
table {%
0 18.75
1 7.97872340425532
2 10.2564102564103
3 5.61797752808989
4 8.43170320404722
5 6.89238210399033
6 6.27431906614786
7 9.32519498134961
};
\addlegendentry{iclr}
\addplot [semithick, crimson2143940, mark=*, mark size=3, mark options={solid}]
table {%
0 7.25075528700906
1 6.94444444444444
2 7.0855614973262
3 6.21632822213013
4 8.2445894881484
5 9.46330777656079
6 9.24558587479936
};
\addlegendentry{neurips}
\end{axis}

\end{tikzpicture}
        \caption{Files and folders which indicate Python packaging adoption over time.}\label{fig:packaging}
    \end{figure}

    Packaging as described in section~\ref{sec:packaging} is an elegant way to improve code reusability.
    This section discusses our estimates of Python code packaging at major \ac{ml} conferences over time.
    Figure~\ref{fig:packaging} illustrates that the numbers stagnate roughly between 20\% and 40\% for \texttt{setup.py} files.
    \Ac{pep}-518 recommends the use of \texttt{pyproject.toml} files over \texttt{setup.py} files since 2016~\citep{Cannon2016pep518}.
    The \ac{pep} also outlines the case against \texttt{setup.py}.
    The trend for \texttt{pyproject.toml} adoption is upwards, which is encouraging.
    The \texttt{setup.cfg} is a configuration file for setuptools, which is used to package Python code.
    Its use is falling, presumably because people are migrating to \texttt{pyproject.toml} files, as recommended.

    We also tracked \texttt{hatch.toml} files.
    In supplementary Figure~\ref{fig:packaging_emerging}, we see a small upward trend for its use, but it's size is insignificant.
    Overall, we see that more projects could package their code.
    This is apparent, especially in comparison to the adoption rates we saw for \ac{mloss} on the right of Figure~\ref{fig:jmlr_stats}.
    The current upward trend for \texttt{pyproject.toml} files is encouraging.
    Packaging is a key component of replicable research, and we should encourage authors to package their code, for example, by including a reference to packaging in the \ac{neurips} code guide~\citep{NeuripsCodeguide}.

    \paragraph{Test and workflow use}
    \begin{figure}
        \centering
\begin{tikzpicture}

\definecolor{crimson2143940}{RGB}{214,39,40}
\definecolor{darkgray176}{RGB}{176,176,176}
\definecolor{darkorange25512714}{RGB}{255,127,14}
\definecolor{forestgreen4416044}{RGB}{44,160,44}
\definecolor{lightgray204}{RGB}{204,204,204}
\definecolor{steelblue31119180}{RGB}{31,119,180}

\begin{axis}[
legend cell align={left},
legend style={
  fill opacity=0.8,
  draw opacity=1,
  text opacity=1,
  at={(0.03,0.97)},
  anchor=north west,
  draw=lightgray204
},
tick align=outside,
tick pos=left,
title={test-folder},
x grid style={darkgray176},
xlabel={conference year},
xmajorgrids,
xmin=-0.35, xmax=7.35,
xtick style={color=black},
xtick={0,1,2,3,4,5,6,7},
xticklabels={18,19,20,21,22,23,24,25},
y grid style={darkgray176},
ylabel={adoption [\%]},
ymajorgrids,
ymin=11.881437443199, ymax=25.4898136928204,
ytick style={color=black}
]
\addplot [semithick, steelblue31119180, mark=*, mark size=3, mark options={solid}]
table {%
0 13.3333333333333
1 13.9175257731959
2 14.9275362318841
3 21.3592233009709
4 19.9460916442049
5 20.8842350203607
6 20.9260588657574
7 24.8712511360194
};
\addlegendentry{icml}
\addplot [semithick, darkorange25512714, mark=*, mark size=3, mark options={solid}]
table {%
0 20.5882352941176
1 19.1176470588235
2 14.5038167938931
3 17.0731707317073
4 18.0672268907563
5 19.1919191919192
6 21.6216216216216
7 18.0365296803653
};
\addlegendentry{aistats}
\addplot [semithick, forestgreen4416044, mark=*, mark size=3, mark options={solid}]
table {%
0 12.5
1 13.8297872340426
2 14.8717948717949
3 14.0449438202247
4 17.2006745362563
5 18.2587666263603
6 23.8326848249027
7 22.3126483553747
};
\addlegendentry{iclr}
\addplot [semithick, crimson2143940, mark=*, mark size=3, mark options={solid}]
table {%
0 14.1993957703927
1 12.6736111111111
2 16.644385026738
3 19.7679237463738
4 19.7870147715562
5 21.5553121577218
6 21.2520064205457
};
\addlegendentry{neurips}
\end{axis}

\end{tikzpicture}
\begin{tikzpicture}

\definecolor{crimson2143940}{RGB}{214,39,40}
\definecolor{darkgray176}{RGB}{176,176,176}
\definecolor{darkorange25512714}{RGB}{255,127,14}
\definecolor{forestgreen4416044}{RGB}{44,160,44}
\definecolor{lightgray204}{RGB}{204,204,204}
\definecolor{steelblue31119180}{RGB}{31,119,180}

\begin{axis}[
legend cell align={left},
legend style={
  fill opacity=0.8,
  draw opacity=1,
  text opacity=1,
  at={(0.03,0.97)},
  anchor=north west,
  draw=lightgray204
},
tick align=outside,
tick pos=left,
title={tox},
x grid style={darkgray176},
xlabel={conference year},
xmajorgrids,
xmin=-0.35, xmax=7.35,
xtick style={color=black},
xtick={0,1,2,3,4,5,6,7},
xticklabels={18,19,20,21,22,23,24,25},
y grid style={darkgray176},
ylabel={adoption [\%]},
ymajorgrids,
ymin=-0.117845117845118, ymax=2.47474747474747,
ytick style={color=black}
]
\addplot [semithick, steelblue31119180, mark=*, mark size=3, mark options={solid}]
table {%
0 0
1 1.03092783505155
2 0.869565217391304
3 1.29449838187702
4 0.988319856244385
5 0.756253635834788
6 0.610193826274228
7 1.0299909118449
};
\addlegendentry{icml}
\addplot [semithick, darkorange25512714, mark=*, mark size=3, mark options={solid}]
table {%
0 0
1 1.47058823529412
2 0
3 0
4 1.26050420168067
5 2.35690235690236
6 1.89189189189189
7 1.36986301369863
};
\addlegendentry{aistats}
\addplot [semithick, forestgreen4416044, mark=*, mark size=3, mark options={solid}]
table {%
0 0
1 0
2 2.05128205128205
3 1.68539325842697
4 0.674536256323777
5 0.725513905683192
6 0.826848249027237
7 1.11902339776195
};
\addlegendentry{iclr}
\addplot [semithick, crimson2143940, mark=*, mark size=3, mark options={solid}]
table {%
0 0.302114803625378
1 1.73611111111111
2 0.601604278074866
3 0.953170327393286
4 0.858811405015459
5 1.13910186199343
6 0.947030497592295
};
\addlegendentry{neurips}
\end{axis}

\end{tikzpicture}
\begin{tikzpicture}

\definecolor{crimson2143940}{RGB}{214,39,40}
\definecolor{darkgray176}{RGB}{176,176,176}
\definecolor{darkorange25512714}{RGB}{255,127,14}
\definecolor{forestgreen4416044}{RGB}{44,160,44}
\definecolor{lightgray204}{RGB}{204,204,204}
\definecolor{steelblue31119180}{RGB}{31,119,180}

\begin{axis}[
legend cell align={left},
legend style={fill opacity=0.8, draw opacity=1, text opacity=1, draw=lightgray204},
tick align=outside,
tick pos=left,
title={noxfile.py},
x grid style={darkgray176},
xlabel={conference year},
xmajorgrids,
xmin=-0.35, xmax=7.35,
xtick style={color=black},
xtick={0,1,2,3,4,5,6,7},
xticklabels={18,19,20,21,22,23,24,25},
y grid style={darkgray176},
ylabel={adoption [\%]},
ymajorgrids,
ymin=-0.147058823529412, ymax=3.08823529411765,
ytick style={color=black}
]
\addplot [semithick, steelblue31119180, mark=*, mark size=3, mark options={solid}]
table {%
0 0
1 0
2 0.144927536231884
3 0.161812297734628
4 0.628930817610063
5 0.232693426410704
6 0.466618808327351
7 0.212056952438655
};
\addlegendentry{icml}
\addplot [semithick, darkorange25512714, mark=*, mark size=3, mark options={solid}]
table {%
0 2.94117647058824
1 0
2 0
3 0
4 0.420168067226891
5 0
6 0.540540540540541
7 0.684931506849315
};
\addlegendentry{aistats}
\addplot [semithick, forestgreen4416044, mark=*, mark size=3, mark options={solid}]
table {%
0 0
1 0.531914893617021
2 0
3 0.561797752808989
4 0
5 0.846432889963724
6 0.535019455252918
7 0.508646998982706
};
\addlegendentry{iclr}
\addplot [semithick, crimson2143940, mark=*, mark size=3, mark options={solid}]
table {%
0 0.302114803625378
1 0.173611111111111
2 0.267379679144385
3 0.372979693327808
4 0.446581930608038
5 0.328587075575027
6 0.401284109149278
};
\addlegendentry{neurips}
\end{axis}

\end{tikzpicture}
\begin{tikzpicture}

\definecolor{crimson2143940}{RGB}{214,39,40}
\definecolor{darkgray176}{RGB}{176,176,176}
\definecolor{darkorange25512714}{RGB}{255,127,14}
\definecolor{forestgreen4416044}{RGB}{44,160,44}
\definecolor{lightgray204}{RGB}{204,204,204}
\definecolor{steelblue31119180}{RGB}{31,119,180}

\begin{axis}[
legend cell align={left},
legend style={
  fill opacity=0.8,
  draw opacity=1,
  text opacity=1,
  at={(0.03,0.97)},
  anchor=north west,
  draw=lightgray204
},
tick align=outside,
tick pos=left,
title={.pre-commit-config.yaml},
x grid style={darkgray176},
xlabel={conference year},
xmajorgrids,
xmin=-0.35, xmax=7.35,
xtick style={color=black},
xtick={0,1,2,3,4,5,6,7},
xticklabels={18,19,20,21,22,23,24,25},
y grid style={darkgray176},
ylabel={adoption [\%]},
ymajorgrids,
ymin=-0.752802181157225, ymax=15.8088458043017,
ytick style={color=black}
]
\addplot [semithick, steelblue31119180, mark=*, mark size=3, mark options={solid}]
table {%
0 2.85714285714286
1 2.57731958762887
2 5.36231884057971
3 9.54692556634304
4 8.08625336927224
5 9.13321698662013
6 12.1679827709978
7 15.0560436231445
};
\addlegendentry{icml}
\addplot [semithick, darkorange25512714, mark=*, mark size=3, mark options={solid}]
table {%
0 2.94117647058824
1 0
2 2.29007633587786
3 7.80487804878049
4 6.72268907563025
5 8.08080808080808
6 8.37837837837838
7 9.36073059360731
};
\addlegendentry{aistats}
\addplot [semithick, forestgreen4416044, mark=*, mark size=3, mark options={solid}]
table {%
0 0
1 5.85106382978723
2 4.1025641025641
3 3.93258426966292
4 5.56492411467116
5 5.56227327690447
6 10.408560311284
7 13.8351983723296
};
\addlegendentry{iclr}
\addplot [semithick, crimson2143940, mark=*, mark size=3, mark options={solid}]
table {%
0 1.81268882175227
1 3.125
2 5.48128342245989
3 7.50103605470369
4 8.72552387495706
5 10.0547645125958
6 12.5361155698234
};
\addlegendentry{neurips}
\end{axis}

\end{tikzpicture}
\begin{tikzpicture}

\definecolor{crimson2143940}{RGB}{214,39,40}
\definecolor{darkgray176}{RGB}{176,176,176}
\definecolor{darkorange25512714}{RGB}{255,127,14}
\definecolor{forestgreen4416044}{RGB}{44,160,44}
\definecolor{lightgray204}{RGB}{204,204,204}
\definecolor{steelblue31119180}{RGB}{31,119,180}

\begin{axis}[
legend cell align={left},
legend style={
  fill opacity=0.8,
  draw opacity=1,
  text opacity=1,
  at={(0.03,0.97)},
  anchor=north west,
  draw=lightgray204
},
tick align=outside,
tick pos=left,
title={.github/workflows},
x grid style={darkgray176},
xlabel={conference year},
xmajorgrids,
xmin=-0.35, xmax=7.35,
xtick style={color=black},
xtick={0,1,2,3,4,5,6,7},
xticklabels={18,19,20,21,22,23,24,25},
y grid style={darkgray176},
ylabel={adoption [\%]},
ymajorgrids,
ymin=-0.57103907906695, ymax=11.9918206604059,
ytick style={color=black}
]
\addplot [semithick, steelblue31119180, mark=*, mark size=3, mark options={solid}]
table {%
0 4.76190476190476
1 3.09278350515464
2 6.23188405797101
3 4.85436893203883
4 7.18778077268643
5 7.38801628853985
6 8.43503230437904
7 11.420781581339
};
\addlegendentry{icml}
\addplot [semithick, darkorange25512714, mark=*, mark size=3, mark options={solid}]
table {%
0 5.88235294117647
1 2.94117647058824
2 3.81679389312977
3 5.36585365853659
4 10.0840336134454
5 7.74410774410774
6 6.21621621621622
7 8.9041095890411
};
\addlegendentry{aistats}
\addplot [semithick, forestgreen4416044, mark=*, mark size=3, mark options={solid}]
table {%
0 0
1 2.12765957446809
2 2.05128205128205
3 1.96629213483146
4 2.86677908937605
5 4.23216444981862
6 6.90661478599222
7 10.6137673787725
};
\addlegendentry{iclr}
\addplot [semithick, crimson2143940, mark=*, mark size=3, mark options={solid}]
table {%
0 1.81268882175227
1 2.60416666666667
2 2.54010695187166
3 4.7244094488189
4 7.90106492614222
5 8.28039430449069
6 9.45425361155698
};
\addlegendentry{neurips}
\end{axis}

\end{tikzpicture}
\begin{tikzpicture}

\definecolor{crimson2143940}{RGB}{214,39,40}
\definecolor{darkgray176}{RGB}{176,176,176}
\definecolor{darkorange25512714}{RGB}{255,127,14}
\definecolor{forestgreen4416044}{RGB}{44,160,44}
\definecolor{lightgray204}{RGB}{204,204,204}
\definecolor{steelblue31119180}{RGB}{31,119,180}

\begin{axis}[
legend cell align={left},
legend style={fill opacity=0.8, draw opacity=1, text opacity=1, draw=lightgray204},
tick align=outside,
tick pos=left,
title={Makefile},
x grid style={darkgray176},
xlabel={conference year},
xmajorgrids,
xmin=-0.35, xmax=7.35,
xtick style={color=black},
xtick={0,1,2,3,4,5,6,7},
xticklabels={18,19,20,21,22,23,24,25},
y grid style={darkgray176},
ylabel={adoption [\%]},
ymajorgrids,
ymin=2.35180412371134, ymax=12.9832474226804,
ytick style={color=black}
]
\addplot [semithick, steelblue31119180, mark=*, mark size=3, mark options={solid}]
table {%
0 4.76190476190476
1 2.83505154639175
2 4.92753623188406
3 4.20711974110032
4 4.67205750224618
5 3.31588132635253
6 5.74300071787509
7 7.33111178430778
};
\addlegendentry{icml}
\addplot [semithick, darkorange25512714, mark=*, mark size=3, mark options={solid}]
table {%
0 5.88235294117647
1 5.88235294117647
2 6.10687022900763
3 2.92682926829268
4 3.78151260504202
5 5.38720538720539
6 4.86486486486486
7 3.65296803652968
};
\addlegendentry{aistats}
\addplot [semithick, forestgreen4416044, mark=*, mark size=3, mark options={solid}]
table {%
0 12.5
1 3.19148936170213
2 3.58974358974359
3 3.37078651685393
4 5.56492411467116
5 3.74848851269649
6 6.22568093385214
7 6.54459138691082
};
\addlegendentry{iclr}
\addplot [semithick, crimson2143940, mark=*, mark size=3, mark options={solid}]
table {%
0 4.83383685800604
1 3.99305555555556
2 3.47593582887701
3 4.47575631993369
4 4.32840948127791
5 6.33077765607886
6 6.51685393258427
};
\addlegendentry{neurips}
\end{axis}

\end{tikzpicture}
        \caption{Systematic test adoption over time. The plot for tox combines counts of \texttt{tox.ini} and \texttt{tox.toml},
            both of which are valid filenames.}\label{fig:tests_ml}
    \end{figure}

    Figure~\ref{fig:tests_ml} illustrates our estimate of the adoption of both methods in parts of the \ac{ml} community over time.
    In the top-left plot in Figure~\ref{fig:tests_ml}, we observe the share of repositories where our crawler found test folders.
    We look at the repository root for \texttt{test} or \texttt{test} as well as within a \texttt{src} or a package folder.
    We estimate that roughly a quarter of repositories have a dedicated test folder.

    Containerized testing requires a specification of dependencies, and a package helps.
    Consequently, our estimates for both are upper bounds.
    The numbers from the Tox and \texttt{noxfile.py} plots in Figure~\ref{fig:tests_ml} allow us to estimate the adoption of containerized testing.
    Since both Tox and Nox run tests in isolated containers.
    We identified automated workflows by checking the presence of the \texttt{.github/workflows} folder and \texttt{.pre-commit-config.yaml} files, which are both used to automate workflows, such as by running tests on every push to a repository.
    Compute-intensive tests are often automated via \texttt{.github/workflows}.
    Linting and formatting checks are less costly and often appear in \texttt{.pre-commit-config.yaml} files.
    Roughly three-quarters of the community work largely without automated testing and workflow automation, even though most projects are team efforts with multiple authors.

    \section{How can we do better?}

    We believe that a step towards improving the community adoption of software best engineering practices could take the form of a checklist for submitters and reviewers, similar to the checklist proposed by~\cite{hoyt2023improving} for the cheminformatics community:

    \begin{itemize}
        \item Does the code have a license file?
        \item Is a README included at the repository root?
        \item Does the project document its dependencies? Does it provide a \texttt{pylock.toml} file?
        If not, are the dependencies documented in the \texttt{pyproject.toml} file?
        \item Does the project facilitate automatic installation by others via packaging? It's never too late to package code! When deadline pressure prevents us from spending the time early, this can still happen between paper acceptance and conference presentation.
        \item Is it possible to run tests automatically? Did we test our external code dependencies by running a containerized set of tests?
        \item Are all pseudorandom number generator seed values fixed?
        \item Did automatic code checkers like flake8 or MyPy (in case of type annotations) report any problems?
        \item Ideally, we would like to have a single file or a command that executes the code required to re-run all experiments from a paper.
    \end{itemize}

    To improve the current situation, we should ask code authors these or similar questions more frequently when reviewing and read the author's answers with an open mindset.

    Sometimes, there are good reasons for a no, which can be an acceptable answer.
    For example, an illustrative notebook for a theory paper does not require a big test machine, since the paper rests on mathematical proofs.
    Furthermore, a mindless test crusade risks triggering authors to withhold code altogether.
    We require a measured approach, which stresses authors' interests and needs.
    Most of the time, authors will re-use their code themselves.
    Moreover, proper software engineering will boost their reach.
    Both arguments can help us to convince authors to adopt best practices.

    Additionally, as a reviewer, if we notice that engineering best practices are not observed, we should share relevant sources from the Python community with the authors, for example, by pointing them to the \ac{neurips} code guide~\citep{NeuripsCodeguide} or to this paper.

    \section{Conclusion}
    
    In \ac{ml} research projects build on each other and teams change over time, motivating software engineering best practices including testing, packaging, and documenting dependencies.
    Their application saves time and effort and also reaps benefits in the long run by preventing bugs, easing on-boarding, and promoting reproducibility.
    Therefore, we advocate for more rigorous software engineering, but with author's interests in mind.
    However, we recognize that there are practical difficulties with applying best practices, including the lack of good incentives, lack of mentorship, competitiveness of the field, and often time pressure.
    Therefore, we are mindful when recommending enforcing software engineering practices that they should be applied only when appropriate, e.g., limiting ourselves to cases where systematic tests are efficient and valuable.
    Forgoing testing, dependency documentation, and packaging initially saves time, which produces a competitive advantage since skipping systematic testing will allow projects to be finished quicker and papers to appear earlier.
    However, our methodology should not be governed by short-term interests.
    As large parts of \ac{ml} research rely on shared code, improving software quality means building a stronger foundation for our work, which is in everyone's best interest.

    \section*{Acknowledgements}
    Research was supported by the Bundesministerium für Bildung und Forschung
    (BMBF) via its "BNTrAInee" (16DHBK1022) and "WestAI" (01IS22094A) projects
    We would like to thank the NFDI4Chem Consortium for support.

    We used Pgfplots \citep{feuersanger2011manual} as well as tikzplotlib \footnote{https://pypi.org/project/tikzplotlib/} to create all plots in this paper.

    \bibliography{position_software}

\begin{thebibliography}{44}
\providecommand{\natexlab}[1]{#1}
\providecommand{\url}[1]{\texttt{#1}}
\expandafter\ifx\csname urlstyle\endcsname\relax
  \providecommand{\doi}[1]{doi: #1}\else
  \providecommand{\doi}{doi: \begingroup \urlstyle{rm}\Url}\fi

\bibitem[Abadi et~al.(2015)Abadi, Agarwal, Barham, Brevdo, Chen, Citro,
  Corrado, Davis, Dean, Devin, Ghemawat, Goodfellow, Harp, Irving, Isard, Jia,
  Jozefowicz, Kaiser, Kudlur, Levenberg, Man\'{e}, Monga, Moore, Murray, Olah,
  Schuster, Shlens, Steiner, Sutskever, Talwar, Tucker, Vanhoucke, Vasudevan,
  Vi\'{e}gas, Vinyals, Warden, Wattenberg, Wicke, Yu, and
  Zheng]{tensorflow2015-whitepaper}
Mart\'{i}n Abadi, Ashish Agarwal, Paul Barham, Eugene Brevdo, Zhifeng Chen,
  Craig Citro, Greg~S. Corrado, Andy Davis, Jeffrey Dean, Matthieu Devin,
  Sanjay Ghemawat, Ian Goodfellow, Andrew Harp, Geoffrey Irving, Michael Isard,
  Yangqing Jia, Rafal Jozefowicz, Lukasz Kaiser, Manjunath Kudlur, Josh
  Levenberg, Dandelion Man\'{e}, Rajat Monga, Sherry Moore, Derek Murray, Chris
  Olah, Mike Schuster, Jonathon Shlens, Benoit Steiner, Ilya Sutskever, Kunal
  Talwar, Paul Tucker, Vincent Vanhoucke, Vijay Vasudevan, Fernanda Vi\'{e}gas,
  Oriol Vinyals, Pete Warden, Martin Wattenberg, Martin Wicke, Yuan Yu, and
  Xiaoqiang Zheng.
\newblock {TensorFlow}: Large-scale machine learning on heterogeneous systems,
  2015.
\newblock URL \url{https://www.tensorflow.org/}.
\newblock Software available from tensorflow.org.

\bibitem[Bradbury et~al.(2025)Bradbury, Frostig, Hawkins, Johnson, Leary,
  Maclaurin, Necula, Paszke, Vander{P}las, Wanderman-{M}ilne, and
  Zhang]{jax2018github}
James Bradbury, Roy Frostig, Peter Hawkins, Matthew~James Johnson, Chris Leary,
  Dougal Maclaurin, George Necula, Adam Paszke, Jake Vander{P}las, Skye
  Wanderman-{M}ilne, and Qiao Zhang.
\newblock {JAX}: composable transformations of {P}ython+{N}um{P}y programs,
  2025.
\newblock URL \url{http://github.com/jax-ml/jax}.

\bibitem[Breck et~al.(2017)Breck, Cai, Nielsen, Salib, and
  Sculley]{breck2017ml}
Eric Breck, Shanqing Cai, Eric Nielsen, Michael Salib, and D~Sculley.
\newblock The ml test score: A rubric for ml production readiness and technical
  debt reduction.
\newblock In \emph{2017 IEEE international conference on big data (big data)},
  pp.\  1123--1132. IEEE, 2017.

\bibitem[Brett~Cannon(2016)]{Cannon2016pep518}
Donald~Stufft Brett~Cannon, Nathaniel J.~Smith.
\newblock Pep 518 – specifying minimum build system requirements for python
  projects, 2016.
\newblock URL \url{https://peps.python.org/pep-0518/}.

\bibitem[Cannon(2024)]{Cannon2024Pep751}
Brett Cannon.
\newblock Pep 751 – a file format to record python dependencies for
  installation reproducibility, 2024.
\newblock URL \url{https://peps.python.org/pep-0751/}.

\bibitem[Cannon et~al.(2020)Cannon, Ingram, Ganssle, Gedam, Eustace, Kluyver,
  and Chung]{Cannon2020Pep621}
Brett Cannon, Dustin Ingram, Paul Ganssle, Pradyun Gedam, Sébastien Eustace,
  Thomas Kluyver, and Tzu-ping Chung.
\newblock Pep 621 – storing project metadata in pyproject.toml, 2020.
\newblock URL \url{https://peps.python.org/pep-0621/}.

\bibitem[Conda-developers(2025)]{conda2025repo}
Conda-developers.
\newblock Conda documentation, 2025.
\newblock URL \url{https://github.com/conda/conda}.

\bibitem[ecosystem authors(2025)]{chex2025docs}
Jax ecosystem authors.
\newblock Chex documentation, 2025.
\newblock URL \url{https://chex.readthedocs.io/en/latest/}.

\bibitem[Feuers{\"a}nger(2011)]{feuersanger2011manual}
Christian Feuers{\"a}nger.
\newblock Manual for package pgfplots.
\newblock \emph{CTAN}, 17, 2011.

\bibitem[Hager(2021)]{Hager2021pdfx}
Chris Hager.
\newblock Pdfx, 2021.
\newblock URL \url{https://github.com/metachris/pdfx}.

\bibitem[Haibe-Kains et~al.(2020)Haibe-Kains, Adam, Hosny, Khodakarami,
  of~Directors Shraddha Thakkar 35 Kusko Rebecca 36 Sansone Susanna-Assunta 37
  Tong Weida 35 Wolfinger Russ D. 38 Mason Christopher E. 39 Jones Wendell 40
  Dopazo Joaquin 41 Furlanello Cesare~42, Waldron, Wang, McIntosh, Goldenberg,
  Kundaje, et~al.]{haibe2020transparency}
Benjamin Haibe-Kains, George~Alexandru Adam, Ahmed Hosny, Farnoosh Khodakarami,
  Massive Analysis Quality Control (MAQC) Society~Board of~Directors Shraddha
  Thakkar 35 Kusko Rebecca 36 Sansone Susanna-Assunta 37 Tong Weida 35
  Wolfinger Russ D. 38 Mason Christopher E. 39 Jones Wendell 40 Dopazo Joaquin
  41 Furlanello Cesare~42, Levi Waldron, Bo~Wang, Chris McIntosh, Anna
  Goldenberg, Anshul Kundaje, et~al.
\newblock Transparency and reproducibility in artificial intelligence.
\newblock \emph{Nature}, 586\penalty0 (7829):\penalty0 E14--E16, 2020.

\bibitem[Hatch-developers(2025)]{hatch2025docs}
Hatch-developers.
\newblock Hatch documentation, 2025.
\newblock URL \url{https://hatch.pypa.io/latest/}.

\bibitem[Heil et~al.(2021)Heil, Hoffman, Markowetz, Lee, Greene, and
  Hicks]{heil2021reproducibility}
Benjamin~J Heil, Michael~M Hoffman, Florian Markowetz, Su-In Lee, Casey~S
  Greene, and Stephanie~C Hicks.
\newblock Reproducibility standards for machine learning in the life sciences.
\newblock \emph{Nature Methods}, 18\penalty0 (10):\penalty0 1132--1135, 2021.

\bibitem[Hoyt et~al.(2023)Hoyt, Zdrazil, Guha, Jeliazkova, Martinez-Mayorga,
  and Nittinger]{hoyt2023improving}
Charles~Tapley Hoyt, Barbara Zdrazil, Rajarshi Guha, Nina Jeliazkova, Karina
  Martinez-Mayorga, and Eva Nittinger.
\newblock Improving reproducibility and reusability in the journal of
  cheminformatics.
\newblock \emph{Journal of Cheminformatics}, 15\penalty0 (1):\penalty0 62,
  2023.

\bibitem[Hutson(2018)]{hutson2018artificial}
Matthew Hutson.
\newblock Artificial intelligence faces reproducibility crisis.
\newblock \emph{Science}, 359, 2018.

\bibitem[Johanson \& Hasselbring(2018)Johanson and
  Hasselbring]{johanson2018software}
Arne Johanson and Wilhelm Hasselbring.
\newblock Software engineering for computational science: Past, present,
  future.
\newblock \emph{Computing in Science \& Engineering}, 20\penalty0 (2):\penalty0
  90--109, 2018.

\bibitem[List et~al.(2017)List, Ebert, and Albrecht]{Lost2017TenUsable}
Markus List, Peter Ebert, and Felipe Albrecht.
\newblock Ten simple rules for developing usable software in computational
  biology.
\newblock \emph{PLoS Comput. Biol.}, 13\penalty0 (1), 2017.
\newblock URL \url{https://doi.org/10.1371/journal.pcbi.1005265}.

\bibitem[mypy developers(2025)]{mypy-team2025docs}
mypy developers.
\newblock mypy, 2025.
\newblock URL \url{https://mypy-lang.org/index.html}.
\newblock Accessed: 2025-07-30.

\bibitem[Paszke et~al.(2017)Paszke, Gross, Chintala, Chanan, Yang, DeVito, Lin,
  Desmaison, Antiga, and Lerer]{paszke2017automatic}
Adam Paszke, Sam Gross, Soumith Chintala, Gregory Chanan, Edward Yang, Zachary
  DeVito, Zeming Lin, Alban Desmaison, Luca Antiga, and Adam Lerer.
\newblock Automatic differentiation in pytorch.
\newblock In \emph{NIPS 2017 Autodiff Workshop}, 2017.

\bibitem[P{\'{e}}rez{-}Riverol et~al.(2016)P{\'{e}}rez{-}Riverol, Gatto, Wang,
  Sachsenberg, Uszkoreit, da~Veiga~Leprevost, Fufezan, Ternent, Eglen, Katz,
  Pollard, Konovalov, Flight, Blin, and Vizca{\'{\i}}no]{Yasset2016TenGit}
Yasset P{\'{e}}rez{-}Riverol, Laurent Gatto, Rui Wang, Timo Sachsenberg, Julian
  Uszkoreit, Felipe da~Veiga~Leprevost, Christian Fufezan, Tobias Ternent,
  Stephen~J. Eglen, Daniel~S. Katz, Tom~J. Pollard, Alexander Konovalov,
  Robert~M. Flight, Kai Blin, and Juan~Antonio Vizca{\'{\i}}no.
\newblock Ten simple rules for taking advantage of git and github.
\newblock \emph{PLoS Comput. Biol.}, 12\penalty0 (7), 2016.
\newblock URL \url{https://doi.org/10.1371/journal.pcbi.1004947}.

\bibitem[Pineau et~al.(2021)Pineau, Vincent-Lamarre, Sinha, Larivi{\`e}re,
  Beygelzimer, d'Alch{\'e} Buc, Fox, and Larochelle]{pineau2021improving}
Joelle Pineau, Philippe Vincent-Lamarre, Koustuv Sinha, Vincent Larivi{\`e}re,
  Alina Beygelzimer, Florence d'Alch{\'e} Buc, Emily Fox, and Hugo Larochelle.
\newblock Improving reproducibility in machine learning research (a report from
  the neurips 2019 reproducibility program).
\newblock \emph{Journal of machine learning research}, 22\penalty0
  (164):\penalty0 1--20, 2021.

\bibitem[Pip-developers(2025)]{pip2025docs}
Pip-developers.
\newblock Requirements file format - pip documentation v25.0, 2025.
\newblock URL
  \url{https://pip.pypa.io/en/stable/reference/requirements-file-format/}.

\bibitem[Popper(2005)]{popper2005logic}
Karl Popper.
\newblock \emph{The logic of scientific discovery}.
\newblock Routledge, 2005.

\bibitem[Princeton-AI-Lab(2025)]{princeton2025reproml}
Princeton-AI-Lab.
\newblock Ml reproducibility challenge, 2025.
\newblock URL \url{https://reproml.org/}.
\newblock Accessed: 2025-06-30.

\bibitem[Prlic \& Procter(2012)Prlic and Procter]{Andreas2012TenSoftware}
Andreas Prlic and James~B. Procter.
\newblock Ten simple rules for the open development of scientific software.
\newblock \emph{PLoS Comput. Biol.}, 8\penalty0 (12), 2012.
\newblock URL \url{https://doi.org/10.1371/journal.pcbi.1002802}.

\bibitem[Pytest-developers(2025)]{pytest2025docs}
Pytest-developers.
\newblock Pytest documentation, 2025.
\newblock URL \url{https://docs.pytest.org/en/stable/}.

\bibitem[Python-developers(2025)]{unittest2025pythondocs}
Python-developers.
\newblock unittest - unit testing framework, 2025.
\newblock URL \url{https://docs.python.org/3/library/unittest.html}.

\bibitem[Python-Packaging-Authority(2025)]{packaging2025python}
Python-Packaging-Authority.
\newblock Packaging python-projects, 2025.
\newblock URL \url{https://packaging.python.org/tutorials/packaging-projects/}.

\bibitem[PyTorch-Contributors(2024)]{PyTorch2024randomness}
PyTorch-Contributors.
\newblock Reproducibility, 2024.
\newblock URL \url{https://pytorch.org/docs/stable/notes/randomness.html}.

\bibitem[Raff(2019)]{raff2019step}
Edward Raff.
\newblock A step toward quantifying independently reproducible machine learning
  research.
\newblock \emph{Advances in Neural Information Processing Systems}, 32, 2019.

\bibitem[Raff(2021)]{raff2021research}
Edward Raff.
\newblock Research reproducibility as a survival analysis.
\newblock In \emph{Proceedings of the AAAI Conference on Artificial
  Intelligence}, volume~35, pp.\  469--478, 2021.
\newblock URL \url{https://doi.org/10.1609/aaai.v35i1.16124}.

\bibitem[Raff(2023)]{raff2023does}
Edward Raff.
\newblock Does the market of citations reward reproducible work?
\newblock In \emph{Proceedings of the 2023 ACM Conference on Reproducibility
  and Replicability}, pp.\  89--96, 2023.

\bibitem[Raff \& Farris(2023)Raff and Farris]{raff2023siren}
Edward Raff and Andrew~L Farris.
\newblock A siren song of open source reproducibility, examples from machine
  learning.
\newblock In \emph{Proceedings of the 2023 ACM Conference on Reproducibility
  and Replicability}, pp.\  115--120, 2023.

\bibitem[Richardson(2023)]{richardson2023soupdocs}
Leonard Richardson.
\newblock Beautiful soup documentation, 2023.
\newblock URL \url{https://www.crummy.com/software/BeautifulSoup/bs4/doc/}.

\bibitem[Rossum et~al.(2014)Rossum, Lehtosalo, and Langa]{vanRossum2014Pep484}
Guido~van Rossum, Jukka Lehtosalo, and Łukasz Langa.
\newblock Pep 484 – type hints, 2014.
\newblock URL \url{https://peps.python.org/pep-0484/}.

\bibitem[Sandve et~al.(2013)Sandve, Nekrutenko, Taylor, and
  Hovig]{Kjetil2013TenReproducible}
Geir~Kjetil Sandve, Anton Nekrutenko, James Taylor, and Eivind Hovig.
\newblock Ten simple rules for reproducible computational research.
\newblock \emph{PLoS Comput. Biol.}, 9\penalty0 (10), 2013.
\newblock URL \url{https://doi.org/10.1371/journal.pcbi.1003285}.

\bibitem[Setuptools-Team(2025)]{setuptools2025docs}
Setuptools-Team.
\newblock Setuptools documentation, 2025.
\newblock URL \url{https://setuptools.pypa.io/en/stable/}.

\bibitem[Smith et~al.(2015)Smith, Kluyver, and Coghlan]{Smith2015pep517}
Nathaniel~J. Smith, Thomas Kluyver, and Alyssa Coghlan.
\newblock Pep 517 – a build-system independent format for source trees, 2015.
\newblock URL \url{https://peps.python.org/pep-0517/}.

\bibitem[Sonnenburg et~al.(2007)Sonnenburg, Braun, Ong, Bengio, Bottou, Holmes,
  LeCun, Müller, Pereira, Rasmussen, et~al.]{sonnenburg2007need}
Sören Sonnenburg, Mikio~L Braun, Cheng~Soon Ong, Samy Bengio, Leon Bottou,
  Geoffrey Holmes, Yann LeCun, Klaus-Robert Müller, Fernando Pereira,
  Carl~Edward Rasmussen, et~al.
\newblock The need for open source software in machine learning.
\newblock \emph{Journal of Machine Learning Research}, 8\penalty0
  (Oct):\penalty0 2443--2466, 2007.
\newblock URL
  \url{https://www.jmlr.org/papers/volume8/sonnenburg07a/sonnenburg07a.pdf}.

\bibitem[Sphinx-developers(2025)]{sphinx2025docs}
Sphinx-developers.
\newblock Sphinx documentation, 2025.
\newblock URL \url{https://www.sphinx-doc.org}.

\bibitem[Stojnic et~al.(2020)Stojnic, Taylor, Pati, Stöter, andShagun Sodhani,
  Husain, Chaudhary, and Jain]{NeuripsCodeguide}
Robert Stojnic, Ross Taylor, Sarthak Pati, Fabian-Robert Stöter, Viktor~Kerkez
  andShagun Sodhani, Hamel Husain, Amit Chaudhary, and Rishabh Jain.
\newblock Tips for releasing research code in machine learning (with official
  neurips 2020 recommendations), 2020.
\newblock URL \url{https://github.com/paperswithcode/releasing-research-code}.
\newblock Accessed: 2025-01-28.

\bibitem[Tatman et~al.(2018)Tatman, VanderPlas, and Dane]{tatman2018practical}
Rachael Tatman, Jake VanderPlas, and Sohier Dane.
\newblock A practical taxonomy of reproducibility for machine learning
  research.
\newblock In \emph{2nd Reproducibility in Machine Learning Workshop at ICML
  2018, Stockholm, Sweden.}, 2018.

\bibitem[van Rossum et~al.(2001)van Rossum, Warsaw, and Coghlan]{pep8}
Guido van Rossum, Barry Warsaw, and Nick Coghlan.
\newblock Pep 8 -- style guide for python code, 2001.
\newblock URL \url{https://www.python.org/dev/peps/pep-0008/}.

\bibitem[Ziemann et~al.(2023)Ziemann, Poulain, and Bora]{ziemann2023five}
Mark Ziemann, Pierre Poulain, and Anusuiya Bora.
\newblock The five pillars of computational reproducibility: bioinformatics and
  beyond.
\newblock \emph{Briefings in Bioinformatics}, 24\penalty0 (6), 2023.

\end{thebibliography}
    \bibliographystyle{tmlr}


    \appendix
    \onecolumn

    \section{Appendix}

    \section*{Impact Statement}
    This work exhibits a critical gap in the adoption of software practices to improve reproducibility in \ac{ml} research and proposes actionable recommendations.
    Our findings aim to call for action in the \ac{ml} community to promote reproducibility and long-term scientific integrity through these standard software practices.

    \printacronyms

    \begin{figure}
\begin{tikzpicture}

\definecolor{crimson2143940}{RGB}{214,39,40}
\definecolor{darkgray176}{RGB}{176,176,176}
\definecolor{darkorange25512714}{RGB}{255,127,14}
\definecolor{forestgreen4416044}{RGB}{44,160,44}
\definecolor{lightgray204}{RGB}{204,204,204}
\definecolor{steelblue31119180}{RGB}{31,119,180}

\begin{axis}[
legend cell align={left},
legend style={fill opacity=0.8, draw opacity=1, text opacity=1, draw=lightgray204},
tick align=outside,
tick pos=left,
title={Pipfile.lock},
x grid style={darkgray176},
xlabel={conference year},
xmajorgrids,
xmin=-0.35, xmax=7.35,
xtick style={color=black},
xtick={0,1,2,3,4,5,6,7},
xticklabels={18,19,20,21,22,23,24,25},
y grid style={darkgray176},
ylabel={adoption [\%]},
ymajorgrids,
ymin=-0.0256410256410256, ymax=0.53,
ytick style={color=black},
scaled y ticks=base 10:1
]
\addplot [semithick, steelblue31119180, mark=*, mark size=3, mark options={solid}]
table {%
0 0
1 0.257731958762887
2 0
3 0
4 0.0898472596585804
5 0.058173356602676
6 0.107681263460158
7 0.0605877006967586
};
\addlegendentry{icml}
\addplot [semithick, darkorange25512714, mark=*, mark size=3, mark options={solid}]
table {%
0 0
1 0
2 0
3 0
4 0
5 0
6 0
7 0
};
\addlegendentry{aistats}
\addplot [semithick, forestgreen4416044, mark=*, mark size=3, mark options={solid}]
table {%
0 0
1 0
2 0.512820512820513
3 0
4 0.168634064080944
5 0
6 0.0486381322957198
7 0.203458799593082
};
\addlegendentry{iclr}
\addplot [semithick, crimson2143940, mark=*, mark size=3, mark options={solid}]
table {%
0 0
1 0
2 0.200534759358289
3 0.165768752590137
4 0.20611473720371
5 0.153340635268346
6 0.14446227929374
};
\addlegendentry{neurips}
\end{axis}

\end{tikzpicture}
\begin{tikzpicture}

\definecolor{crimson2143940}{RGB}{214,39,40}
\definecolor{darkgray176}{RGB}{176,176,176}
\definecolor{darkorange25512714}{RGB}{255,127,14}
\definecolor{forestgreen4416044}{RGB}{44,160,44}
\definecolor{lightgray204}{RGB}{204,204,204}
\definecolor{steelblue31119180}{RGB}{31,119,180}

\begin{axis}[
legend cell align={left},
legend style={fill opacity=0.8, draw opacity=1, text opacity=1, draw=lightgray204},
tick align=outside,
tick pos=left,
title={pixi.lock},
x grid style={darkgray176},
xlabel={conference year},
xmajorgrids,
xmin=-0.35, xmax=7.35,
xtick style={color=black},
xtick={0,1,2,3,4,5,6,7},
xticklabels={18,19,20,21,22,23,24,25},
y grid style={darkgray176},
ylabel={adoption [\%]},
ymajorgrids,
ymin=-0.0120918984280532, ymax=0.53,
ytick style={color=black},
scaled y ticks=base 10:1
]
\addplot [semithick, steelblue31119180, mark=*, mark size=3, mark options={solid}]
table {%
0 0
1 0
2 0.144927536231884
3 0
4 0.0898472596585804
5 0
6 0
7 0.0908815510451378
};
\addlegendentry{icml}
\addplot [semithick, darkorange25512714, mark=*, mark size=3, mark options={solid}]
table {%
0 0
1 0
2 0
3 0
4 0
5 0
6 0
7 0
};
\addlegendentry{aistats}
\addplot [semithick, forestgreen4416044, mark=*, mark size=3, mark options={solid}]
table {%
0 0
1 0
2 0
3 0
4 0.168634064080944
5 0.241837968561064
6 0.0486381322957198
7 0.0678195998643608
};
\addlegendentry{iclr}
\addplot [semithick, crimson2143940, mark=*, mark size=3, mark options={solid}]
table {%
0 0
1 0
2 0.0668449197860962
3 0
4 0
5 0.0219058050383352
6 0.0481540930979133
};
\addlegendentry{neurips}
\end{axis}

\end{tikzpicture}
        \caption{Rare requirements documentation \texttt{lock}-files}\label{fig:lock_other}
    \end{figure}

    \begin{figure}
\begin{tikzpicture}

\definecolor{crimson2143940}{RGB}{214,39,40}
\definecolor{darkgray176}{RGB}{176,176,176}
\definecolor{darkorange25512714}{RGB}{255,127,14}
\definecolor{forestgreen4416044}{RGB}{44,160,44}
\definecolor{lightgray204}{RGB}{204,204,204}
\definecolor{steelblue31119180}{RGB}{31,119,180}

\begin{axis}[
legend cell align={left},
legend style={fill opacity=0.8, draw opacity=1, text opacity=1, draw=lightgray204},
tick align=outside,
tick pos=left,
title={pixi.toml},
x grid style={darkgray176},
xlabel={conference year},
xmajorgrids,
xmin=-0.35, xmax=7.35,
xtick style={color=black},
xtick={0,1,2,3,4,5,6,7},
xticklabels={18,19,20,21,22,23,24,25},
y grid style={darkgray176},
ylabel={adoption [\%]},
ymajorgrids,
ymin=-0.00809061488673139, ymax=0.169902912621359,
ytick style={color=black},
scaled y ticks=base 10:2
]
\addplot [semithick, steelblue31119180, mark=*, mark size=3, mark options={solid}]
table {%
0 0
1 0
2 0.144927536231884
3 0.161812297734628
4 0.0898472596585804
5 0
6 0
7 0.0605877006967586
};
\addlegendentry{icml}
\addplot [semithick, darkorange25512714, mark=*, mark size=3, mark options={solid}]
table {%
0 0
1 0
2 0
3 0
4 0
5 0
6 0
7 0
};
\addlegendentry{aistats}
\addplot [semithick, forestgreen4416044, mark=*, mark size=3, mark options={solid}]
table {%
0 0
1 0
2 0
3 0
4 0
5 0
6 0
7 0
};
\addlegendentry{iclr}
\addplot [semithick, crimson2143940, mark=*, mark size=3, mark options={solid}]
table {%
0 0
1 0
2 0.0668449197860962
3 0
4 0
5 0.0219058050383352
6 0.0160513643659711
};
\addlegendentry{neurips}
\end{axis}

\end{tikzpicture}
\begin{tikzpicture}

\definecolor{crimson2143940}{RGB}{214,39,40}
\definecolor{darkgray176}{RGB}{176,176,176}
\definecolor{darkorange25512714}{RGB}{255,127,14}
\definecolor{forestgreen4416044}{RGB}{44,160,44}
\definecolor{lightgray204}{RGB}{204,204,204}
\definecolor{steelblue31119180}{RGB}{31,119,180}

\begin{axis}[
legend cell align={left},
legend style={
  fill opacity=0.8,
  draw opacity=1,
  text opacity=1,
  at={(0.03,0.97)},
  anchor=north west,
  draw=lightgray204
},
tick align=outside,
tick pos=left,
title={hatch.toml},
x grid style={darkgray176},
xlabel={conference year},
xmajorgrids,
xmin=-0.35, xmax=7.35,
xtick style={color=black},
xtick={0,1,2,3,4,5,6,7},
xticklabels={18,19,20,21,22,23,24,25},
y grid style={darkgray176},
ylabel={adoption [\%]},
ymajorgrids,
ymin=-0.00240770465489567, ymax=0.050561797752809,
ytick style={color=black},
scaled y ticks=base 10:2
]
\addplot [semithick, steelblue31119180, mark=*, mark size=3, mark options={solid}]
table {%
0 0
1 0
2 0
3 0
4 0
5 0
6 0
7 0.0302938503483793
};
\addlegendentry{icml}
\addplot [semithick, darkorange25512714, mark=*, mark size=3, mark options={solid}]
table {%
0 0
1 0
2 0
3 0
4 0
5 0
6 0
7 0
};
\addlegendentry{aistats}
\addplot [semithick, forestgreen4416044, mark=*, mark size=3, mark options={solid}]
table {%
0 0
1 0
2 0
3 0
4 0
5 0
6 0
7 0
};
\addlegendentry{iclr}
\addplot [semithick, crimson2143940, mark=*, mark size=3, mark options={solid}]
table {%
0 0
1 0
2 0
3 0
4 0
5 0
6 0.0481540930979133
};
\addlegendentry{neurips}
\end{axis}

\end{tikzpicture}
        \caption{Adoption of potentially emerging packaging tools.}
        \label{fig:packaging_emerging}
    \end{figure}
\end{document}